\documentclass[pdflatex,sn-mathphys-num]{sn-jnl}

\usepackage{amsmath}
\usepackage{graphicx}%
\usepackage{multirow}%
\usepackage{amsmath,amssymb,amsfonts}%
\usepackage{amsthm}%
\usepackage{mathrsfs}%
\usepackage[title]{appendix}%
\usepackage{xcolor}%
\usepackage{textcomp}%
\usepackage{manyfoot}%
\usepackage{booktabs}%
\usepackage{algorithm}%
\usepackage{algorithmicx}%
\usepackage{algpseudocode}%
\usepackage{listings}%
\usepackage{tikz}
\usepackage{rotating, tabularx, booktabs, array}
\newcolumntype{C}{>{\centering\arraybackslash}X}

\usepackage{enumitem}
\usepackage{placeins}

\theoremstyle{thmstyleone}%
\theoremstyle{thmstyletwo}%

\theoremstyle{thmstylethree}%

\begin{document}

\title[Article Title]{Carbon Reduction Potential and Sensitivity Analysis of Rural Integrated Energy System with Carbon Trading and Coordinated Electric–Thermal Demand Response}


\author[1]{\fnm{Xuxin} \sur{Yang}}\email{yangxuxin@sjtu.edu.cn}

\author[1]{\fnm{Xue} \sur{Yuan}}\email{yxyuanxue@sjtu.edu.cn}

\author*[1]{\fnm{Donghan} \sur{Feng}}\email{seed@sjtu.edu.cn}

\author[2]{\fnm{Siru} \sur{Chen}}\email{chensiru@sjtu.edu.cn}

\author[1]{\fnm{Yuanhao} \sur{Feng}}\email{fyh1999@sjtu.edu.cn}

\affil*[1]{\orgdiv{School of Electrical Engineering}, \orgname{Shanghai Jiao Tong University}, \orgaddress{\street{800 Dongchuan Road}, \city{Shanghai}, \postcode{200240}, \state{Shanghai}, \country{China}}}

\affil[2]{\orgdiv{College of Smart Energy}, \orgname{Shanghai Jiao Tong University}, \orgaddress{\street{800 Dongchuan Road}, \city{Shanghai}, \postcode{200240}, \state{Shanghai}, \country{China}}}


\abstract{Constructing clean and low-carbon rural integrated energy system (RIES) is a fundamental requirement for supporting China’s rural modernization and new-type urbanization. Existing research on RIES decarbonization primarily focuses on the optimal low-carbon operation of system-level energy devices at the macro level, while the synergistic carbon-reduction effects of demand-side flexible loads and external carbon trading mechanisms have not been fully explored. Meanwhile, at the micro level, the carbon sensitivity of device parameters and their potential contribution to emission reduction remain insufficiently investigated. To address these gaps, this study integrates macro- and micro-level analyses. At the macro level, a multi-energy-coupled low-carbon optimal operation framework is developed, incorporating coordinated electric–thermal demand response (DR) and carbon trading. At the micro level, a carbon emission model for RIES components is established, and sensitivity analysis is conducted on 28 carbon-related parameters to identify highly sensitive determinants of emission reduction. Case studies based on typical operation data from a rural region in northern China demonstrate that coordinated electric–thermal DR and carbon trading can achieve maximum carbon-reduction potential. Furthermore, the identified high-sensitivity parameters provide essential theoretical guidance for enhancing the decarbonization potential of RIES.}




\maketitle

\section{Introduction}\label{sec1}
\noindent\hspace{1em}
China is recognized as having the second-largest rural population in the world, with approximately 550 million rural residents. From 2000 to 2017, carbon emissions from household energy consumption in rural China were observed to increase at an average annual rate of 4.8\%, accounting for 42\% of the nation’s residential energy-related carbon emissions [1-3]. In addition, the green and low-carbon transformation of the rural energy system has been regarded as not only a key pathway for achieving China’s “carbon peaking–carbon neutrality” strategic goals, but also an essential foundation for rural modernization and new-type urbanization [4-6]. In recent years, notable progress has been achieved in the deployment and application of integrated energy system (IES) in rural areas, driven by the rural revitalization strategy [7-10]. Supported by renewable resources such as biogas, crop straw, and biomass waste, rural integrated energy system (RIES) have been demonstrated to provide diverse energy conversion pathways, flexible system configurations, and high energy-use efficiency, thereby offering a promising direction for clean and low-carbon rural energy development [11-15]. Therefore, the low-carbon operation mechanisms and carbon-reduction potential of RIES are considered to merit systematic investigation due to their significant theoretical and practical relevance.\\

Relatively few studies have investigated carbon reduction in RIES. According to Ref.[16], approximately 40\% of RIES in China are located in Shandong, Heilongjiang, Shanxi, and Hebei Provinces combined. The study also analyzed the carbon-reduction potential of deploying electrified equipment, such as air-source heat pump (HP), in rural areas. The results indicated that, through the coordinated deployment of RIES and electrified terminal devices, Carbon emissions in the rural residential sector could be reduced by approximately 95\% by 2060 compared with the 2014 level. Ref.[17] investigated rural areas in Lankao County, China, and developed a multi-energy complementary RIES model integrating Power-to-Gas (P2G) and Waste Incineration (WI) technologies, considering abundant local biomass resources such as crop straw and household waste, as well as rooftop photovoltaic (PV) and distributed wind power (WP). Multiple objectives including risk, revenue, and carbon emissions were jointly considered, and a Nash-bargaining-based multi-agent benefit allocation strategy was adopted to achieve optimal electricity–carbon–electricity cycling and efficient aggregation of renewable energy within the RIES. Ref.[18] proposed a low-carbon optimal planning method for rural industrial parks, modeling the zero-carbon and carbon-reduction benefits of biomass energy on the generation side, and establishing carbon emission models for rural irrigation systems and straw-combustion processes. With the objective of minimizing annualized total cost and incorporating carbon-reduction targets into the optimization framework, the results demonstrated that system annual carbon emissions could be reduced from 37,533.98 tCO$_2$ to 15,834.5 tCO$_2$. Furthermore, Ref.[19] proposed an optimal planning method for RIES integrating biomass energy and solar energy for rural areas. The results showed that, compared with systems adopting biomass energy or solar energy alone, the proposed configuration could reduce annualized total cost by 40.34\% and 28.09\%, decrease carbon emissions by 80.33\% and 67.27\%, and improve energy-use efficiency by 35.33\% and 20.31\%, respectively.\\

In summary, although existing studies have made certain progress in exploring carbon reduction pathways for RIES, several limitations remain:\\

(1) At the macro level: Most studies on RIES concentrate on renewable energy on the power-generation side, focusing on the configuration and optimal scheduling of renewable generation resources. However, the applicability of tiered carbon-trading mechanisms in RIES and the potential for coordinated low-carbon and economic optimal scheduling have been largely overlooked [20-22]. In addition, limited attention has been paid to the synergistic effects between demand-side flexible load responses and external carbon-market mechanisms [23,24], and the carbon-reduction potential under different operating scenarios has not been sufficiently investigated [25]. Existing studies on DR can be broadly divided into two categories: electricity-and-heat DR driven by electricity price signals [26-28], and low-carbon DR based on dynamic carbon-emission factors(CEF) [29-31].\\

(2) At the micro level: Current research lacks quantitative sensitivity analysis between key equipment parameters in RIES and carbon emissions, making it difficult to clarify and quantify how parameter variations at the equipment level affect overall system-level carbon-reduction performance.\\

To address the above challenges, research pathways for analyzing the carbon-reduction potential of RIES can be informed by studies conducted at both macro and micro levels.\\

(1) At the macro level: Ref.[32] incorporated electric–thermal DR and a carbon-trading mechanism into the optimal scheduling model of an IES to achieve low-carbon and cost-efficient operation. The results indicated that introducing a tiered carbon-trading mechanism reduced the total system cost by 5.9\%, and further integration of electric–thermal DR decreased the operating cost by an additional 3.1\%. Moreover, the synergistic interaction between electric–thermal DR and the tiered carbon-trading mechanism enabled flexible load regulation, reduced natural gas consumption, and effectively mitigated overall system carbon emissions. Ref.[33] applied a tiered carbon-trading mechanism to a low-carbon dispatch model for an integrated offshore wind-to-hydrogen and dynamic hydrogen-blending energy system, verifying its feasibility and economic advantages in multi-energy coordinated optimization and carbon-emission reduction. Ref.[34] further proposed a DR method based on a dual-incentive mechanism, wherein dynamic CEF of multi-energy systems were utilized to guide users in actively reducing both carbon-emission quantities and associated costs, thereby achieving significant decarbonization benefits.\\

(2) At the micro level: Ref.[35] performed a carbon-emission sensitivity analysis of multiple high-emission devices within IES to identify critical emission-reduction pathways during system operation, thereby providing valuable insights for operators to design effective carbon-mitigation strategies. The findings assist system operators in optimizing operational schemes based on the carbon-sensitivity characteristics of key devices, effectively lowering carbon emissions while enhancing overall system efficiency and decarbonization performance. Ref.[36] further introduced a quantitative carbon-sensitivity index and applied it to RIES in a less-developed area of China (Jia County as a case study). The study validated the effectiveness of this index in identifying dominant influencing factors, quantifying device-level carbon-emission responses, and guiding low-carbon operational strategies for RIES.\\

To address the above limitations, the research framework is illustrated in Fig. 1, and the main contributions at both the macro and micro levels include:\\
\begin{figure}[H]
\centering
\includegraphics[width=0.8\textwidth]{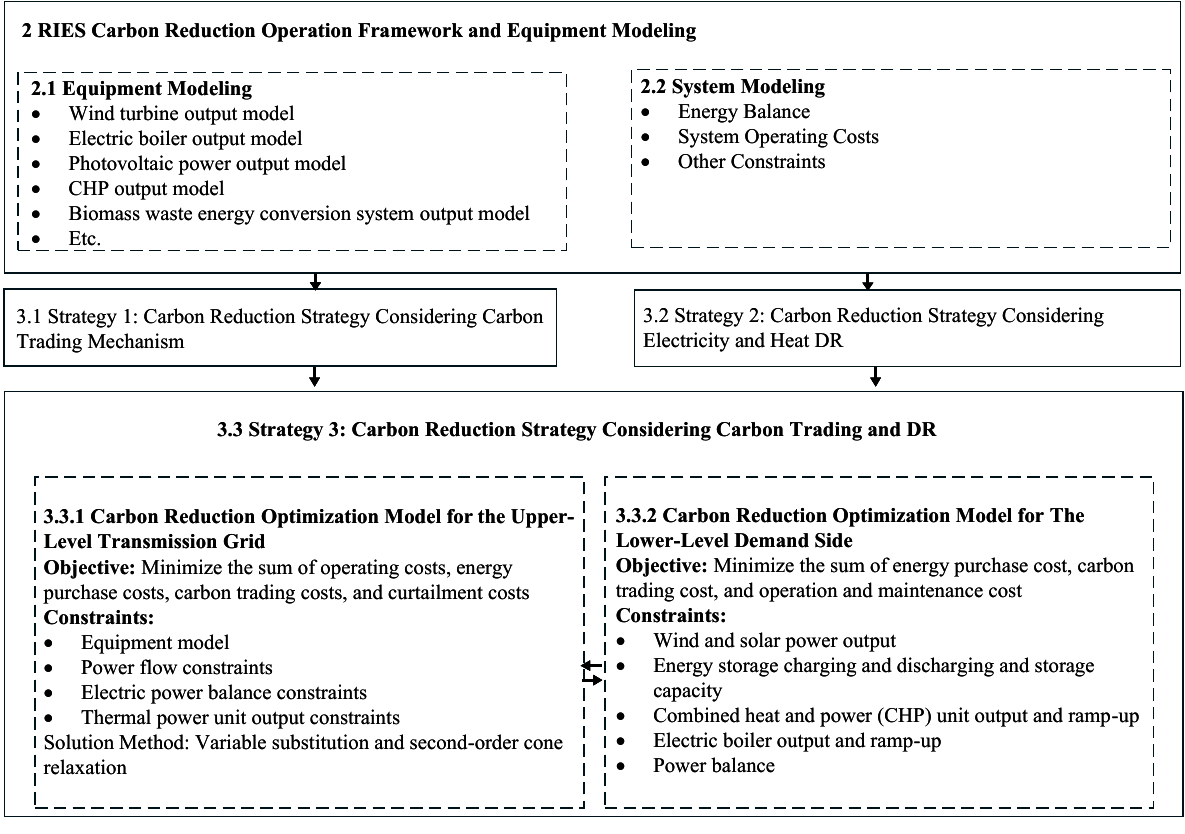}
\caption{Electric-thermal RIES structure}
\label{fig1}
\end{figure}
(1)Development of a low-carbon operational model for RIES: Biogas and crop-straw biomass power generation are integrated on the supply side to achieve efficient coupling and utilization of renewable energy within RIES.\\

(2)Proposal of a coordinated carbon-reduction strategy considering electricity–heat DR and carbon-trading mechanisms: An external carbon-trading mechanism is incorporated on the grid side, while a rural electricity–thermal DR model based on dynamic CEF is established on the load side. The independent and coordinated carbon-reduction effects of both mechanisms are systematically evaluated.\\

(3)Implementation of equipment-level carbon sensitivity analysis and carbon reduction potential assessment: Carbon sensitivity quantification is performed for key equipment parameters at the micro level. Building on the macro-level coordinated optimal operation results, this study proposes targeted carbon-reduction pathways and operational strategies for RIES, culminating in a set of concrete recommendations.\\

\section{RIES carbon reduction operation framework and equipment modeling}\label{sec2}
\noindent\hspace{1em}
The RIES investigated in this study is illustrated in Fig.2, comprising the supply side, the energy network, and the park-level RIES. The energy network primarily consists of the natural gas network and the electricity grid, while the generation side includes conventional carbon-based units represented by gas-fired and coal-fired power plants. At the lower level, the park-scale RIES integrates biomass power units utilizing crop straw and agricultural waste [37,38], distributed PV and wind turbine (WT) units, combined heat and power (CHP) units, electric boiler (EB), biogas-to-grid (B2G), gas turbine (GT), gas boiler (GB), electricity–heat (EH) coupling devices, and energy storage (ES) systems [39,40]. The detailed mathematical models and operational constraints of these components are elaborated in this chapter.

\begin{figure}[h]
\centering
\includegraphics[width=0.8\textwidth]{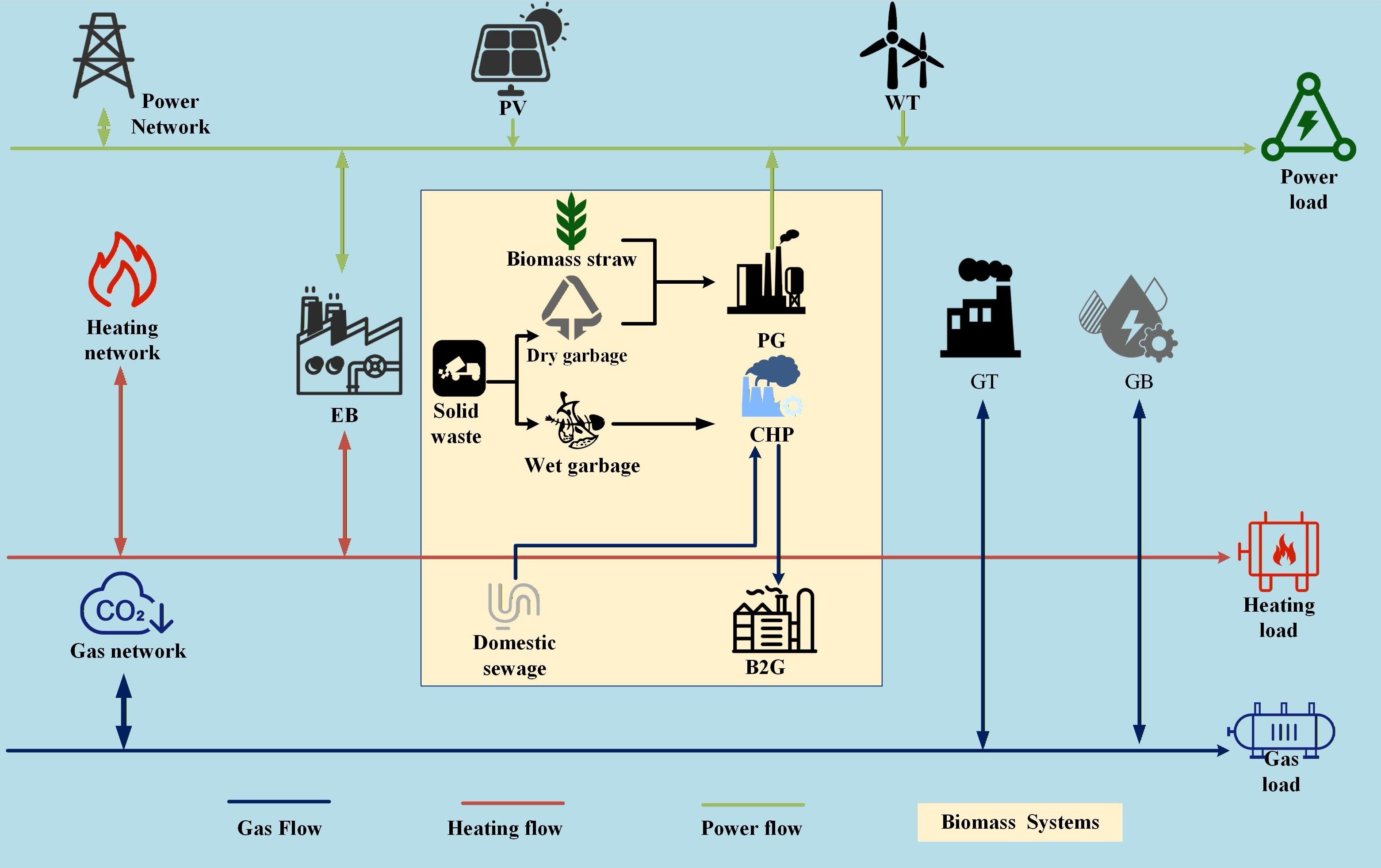}
\caption{Electric and thermal RIES structure}\label{fig1}
\end{figure}
\subsection{RIES equipments modeling}\label{subsec2}
\subsubsection{Model of non-carbon-based energy equipments}\label{subsubsec2}

(1) Model of WT output\\

The relationship between the WT output power and the actual wind speed is shown in Eq.1.
\begin{equation}
\left\{
\begin{aligned}
&P_{\mathrm{wt}}(v) = 0 &\quad& v < v_{\mathrm{in}},\ v > v_{\mathrm{out}} \\
&P_{\mathrm{wt}}(v) = \frac{v - v_{\mathrm{in}}}{v_* - v_{\mathrm{in}}} P_* &\quad& v_{\mathrm{in}} \leq v < v_* \\
&P_{\mathrm{wt}}(v) = P_* &\quad& v_* \leq v < v_{\mathrm{out}}
\end{aligned}
\right.
\label{eq1}
\end{equation}

\noindent Where $P_{\mathrm{wt}}(v)$ denotes the output power of WT (kW); 
$v_{\mathrm{in}}$ and $v_{\mathrm{out}}$ represent the cut-in and cut-out wind speeds (m/s); 
$v^{*}$ is the rated wind speed (m/s); 
$P^{*}$ is the rated power of the turbine (kW).\\
(2) Model of PV output\\

The output characteristics of a PV array are affected by the coupling of multiple factors, including the physical parameters of the components, ambient temperature, and solar irradiance. Existing research shows that solar irradiance exhibits a Beta-like distribution during the diurnal cycle, and its instantaneous output power can be calculated using Eq.2.
\begin{equation}
P_{\mathrm{pv},t}=P_{\mathrm{pv},\max}\frac{G_t}{G_\mathrm{N}}(1+k(T_\mathrm{c}(t)-T_\mathrm{r}))\label{eq2}
\end{equation}

\noindent Where $P_{\mathrm{pv},t}$ denotes the PV power output at time $t$ (kW); 
$P_{\mathrm{pv},\max}$ is the rated PV capacity (kW); 
$G_t$ and $G_\mathrm{N}$ are the actual and nominal solar irradiance (W/m$^2$); 
$k$ is the temperature coefficient; 
$T_\mathrm{c}(t)$ and $T_\mathrm{r}$ represent the cell temperature and the reference temperature (°C).\\
(3) Model of biomass waste energy conversion system\\ 
1) Pyrolysis gasification (PG) system\\

When a PG system is adopted, the organic components in crop straw and municipal solid waste decompose into combustible gases—primarily hydrogen and methane—under high-temperature (600–800 °C) and oxygen-deficient conditions. These gases are subsequently combusted at 900–1000 °C to drive an internal combustion engine for electricity generation. Furthermore, the high-temperature flue gas produced during combustion can be utilized for heating through a waste heat boiler, thereby improving overall energy efficiency. The PG consists of two main sections. The first section involves the pyrolysis gasification process, wherein the feedstock undergoes thermal decomposition to produce a mixture of syngas, tar, and char under controlled temperature and limited oxygen conditions.
\begin{equation}
F_{\mathrm{bio},t}=
\begin{bmatrix}
M_{\mathrm{straw},t}\beta_{\mathrm{straw}} \\
M_{\mathrm{garbage},t}\beta_{\mathrm{garbage}}
\end{bmatrix}^T
\begin{bmatrix}
\beta_{\mathrm{straw},\mathrm{r2f}} \\
\beta_{\mathrm{garbage},\mathrm{r2f}}
\end{bmatrix}\eta_{\mathrm{pf}}\label{eq1}
\end{equation}

\noindent Where $F_{\mathrm{bio},t}$ is the biomass feed flow (kg); 
$M_{\mathrm{straw},t}$ and $M_{\mathrm{garbage},t}$ are the mass of straw and municipal waste (kg); 
$\beta_{\mathrm{straw}}$ and $\beta_{\mathrm{garbage}}$ are their conversion coefficients; 
$\beta_{\mathrm{straw},\mathrm{r2f}}$ and $\beta_{\mathrm{garbage},\mathrm{r2f}}$ are raw-to-fuel ratios; 
$\eta_{\mathrm{pf}}$ is the pyrolysis efficiency.\\
2) Gas production model of biomass waste\\

Synthetic natural gas production from biomass waste is achieved through a two-step process: first, the co-digestion of manure and moist organic waste to produce biogas, and second, the purification and upgrading of this biogas. This entire chain, from storage to processing, necessitates continuous inputs of electrical and thermal energy.\\

Step 1: Model of biogas production from manure and organic waste.\\

Co-digesting manure with organic waste optimizes the carbon-to-nitrogen (C/N) ratio, as the former is nitrogen-rich while the latter is carbon-rich. This balanced ratio enhances microbial activity and anaerobic digestion efficiency, leading to increased biogas yield. In the proposed system, the Sedimentation Tank unit first sediments rural domestic wastewater. The resulting sludge is then co-fermented with moist organic waste, while the clarified effluent is conveyed to the B2G and P2G units for synthetic natural gas synthesis.
\begin{flalign}
\left\{
\begin{array}{l}
F_{\mathrm{st},t}=g_{\mathrm{ST},t}\beta_{\mathrm{ST}}\eta_{AB} \\
M_{\mathrm{sludge},t}=V_{\mathrm{ST},t}\beta_{\mathrm{sludge}}\rho_{\mathrm{sludge}} \\
F_{\mathrm{bg,t}}=(M_{\mathrm{sludge},t}+M_{\mathrm{wet-garbage},t})\beta_{\mathrm{BG}}
\end{array}
\right.
\label{eq1}
\end{flalign}

\noindent Where $F_{\mathrm{st},t}$ is the sludge feed (kg); 
$M_{\mathrm{sludge},t}$ and $V_{\mathrm{ST},t}$ denote the mass and volume of sludge; 
$\rho_{\mathrm{sludge}}$ is the sludge density (kg/m$^3$); 
$\beta_{\mathrm{sludge}}$ and $\beta_{\mathrm{BG}}$ are conversion ratios; 
$\eta_{\mathrm{AB}}$ is the anaerobic digestion efficiency.
The biogas digester requires a constant supply of heat to maintain an optimal internal temperature supporting anaerobic microbial activity.
\begin{equation}
\begin{cases}
P_{\mathrm{bd},t}^{\mathrm{h}}=S_{\mathrm{BD}}\eta_{\mathrm{BD}}\Delta T_t=g_{\mathrm{BD},t}\eta_{\mathrm{EQ}} \\
\eta_{\mathrm{BD}}=\frac{1}{1/\alpha_1+1/\alpha_2+\varphi_1/\theta_1+\varphi_2/\theta_2} & 
\end{cases}\label{eq1}
\end{equation}

\noindent Where $P^{\mathrm{h}}_{\mathrm{bd},t}$ is the heat output of the biogas reactor (kW); 
$S_{\mathrm{BD}}$ is the effective heat exchange area (m$^2$); 
$\eta_{\mathrm{BD}}$ and $\eta_{EQ}$ denote the heat exchange and equipment efficiency; 
$\Delta T_t$ is the temperature difference (°C); 
$\alpha_1$, $\alpha_2$, $\phi_1$, $\phi_2$, $\theta_1$, and $\theta_2$ are empirical heat transfer coefficients.\\

Step 2: Model of biogas upgrading to natural gas.\\

In the B2G unit, sulfur compounds and carbon dioxide are removed from raw biogas through water scrubbing and membrane separation. This purification and upgrading process converts biogas, which typically contains about 60\% methane, into pipeline-quality natural gas with a methane content exceeding 95\%.
\begin{equation}
F_{\mathrm{ng},t}=F_{\mathrm{bg},t}\eta_{\mathrm{B2G}}\label{eq1}
\end{equation}

\noindent Where $F_{\mathrm{ng},t}$ and $F_{\mathrm{bg,t}}$ denote the flow of natural gas and biogas (Nm$^3$); 
$\eta_{\mathrm{B2G}}$ is the conversion efficiency from biogas to natural gas. Eq.4 models the biogas production from mixed organic feedstocks, Eq.5 quantifies the heat requirement of the digester, and Eq.6 describes the biogas upgrading process to natural gas.\\
(4) Model of CHP output\\

This study utilizes biomass-derived syngas as the sole fuel input for the CHP unit. Unlike conventional CHP systems powered by natural gas, the CHP system considered here operates on biomass-generated syngas for CHP production. Consequently, a first-principles, efficiency-based model is employed in place of the empirical quadratic fuel-to-power relationship commonly used for commercial natural-gas-fired CHP units.

\begin{equation}
\left\{
\begin{aligned}
g_{\mathrm{PG},t} &= F_{\mathrm{bio},t}\cdot L_{\mathrm{fuel}} \cdot \eta_{\mathrm{pg}} \\
g_{\mathrm{NG},t} &= F_{\mathrm{ng},t} \cdot L_{\mathrm{ng}}\\
F_{\mathrm{tot},t} &= F_{\mathrm{pg,t}} + F_{\mathrm{ng},t}\\
P_{\mathrm{chp},t}^{\mathrm{e}} &= \eta^{\mathrm{chp,e}} \cdot P_{\mathrm{chp},t}\\
P_{\mathrm{chp},t}^{\mathrm{h}} &= \eta^{\mathrm{chp,h}} \cdot P_{\mathrm{chp},t}\\
\end{aligned}
\right.
\label{eq6}
\end{equation}

\noindent Where $F_{\mathrm{bio},t}$ and $F_{\mathrm{ng},t}$ are the biomass and natural gas inputs (Nm$^3$); 
$L_{\mathrm{fuel}}$ and $L_{\mathrm{ng}}$ are their lower heating values (MJ/Nm$^3$); 
$P^{\mathrm{e}}_{\mathrm{chp},t}$ and $P^{\mathrm{h}}_{\mathrm{chp},t}$ denote the electric and heat output of CHP (kW); 
$\eta_{\mathrm{chp},e}$ and $\eta_{\mathrm{chp},h}$ are the electric and heat efficiency.
(5) Model of EB output\\

The relationship between the heating output power and power consumption of an EB is shown in the following formula:
\begin{equation}
P_{\mathrm{eb},t}^{\mathrm{h}}=\eta^{\mathrm{eb}}P_{\mathrm{eb},t}^{\mathrm{e}}\label{eq6}
\end{equation}
\noindent Where $P^{\mathrm{h}}_{\mathrm{eb},t}$ and $P^{\mathrm{e}}_{\mathrm{eb},t}$ denote the heat output and power input of the EB (kW); 
$\eta_{\mathrm{eb}}$ is the boiler efficiency.

\subsubsection{Model of carbon-based energy equipments}\label{subsubsec2}
(1) Model of GT output\\

The relationship between GT output and natural gas consumption is shown in the following equation:
\begin{equation}
P_{\mathrm{gt},t}^e=\eta^{\mathrm{gt}}G_t^{\mathrm{gt}}\label{eq4}
\end{equation}

\noindent Where $P^{e}_{\mathrm{gt},t}$ is the electrical output of the GT (kW); 
$\eta_{\mathrm{gt}}$ is the turbine efficiency; 
$G_{\mathrm{gt},t}$ is the gas consumption (Nm$^3$).\\
(2) Model of GB output\\

The GB is a heating device that uses natural gas as fuel. The relationship between its heating output and natural gas consumption is shown in Eq.10:
\begin{equation}
\left\{
\begin{aligned}
P_{\mathrm{gb},t} &= G_{t}^{\mathrm{gb}}\\
P_{\mathrm{gb},t}^h&= \eta^{\mathrm{gb,h}} \cdot p_{t}^{\mathrm{gb}}\\
F_{\mathrm{gb},t}^{\mathrm{gas}}&= a^{\mathrm{gb}} (p_{t}^{\mathrm{gb}})^2 + b^{\mathrm{gb}} p_{t}^{\mathrm{gb}} + c^{\mathrm{gb}}\\
E_{t}^{\mathrm{gb}} &= k_{\mathrm{CO_2}}^{\mathrm{gas}} F_{\mathrm{gb},t}
\end{aligned}
\right.
\end{equation}
\noindent Where $P^{h}_{\mathrm{gb},t}$ is the heat output of the gas boiler (kW); 
$\eta_{\mathrm{gb},h}$ is its heat efficiency; 
$F_{\mathrm{gb},t}$ is the gas consumption (Nm$^3$).

\subsubsection{Model of rural residential and agricultural energy}\label{subsubsec2}
\noindent\hspace{1em}
Rural electricity consumption can be categorized into four primary types: residential, small industrial, electric vehicle (EV) charging, and agricultural loads [41-45]. Unlike in urban settings, rural EV typically engage in unidirectional charging without discharging capability. In terms of demand flexibility, residential and agricultural loads are generally rigid, meaning they can only be curtailed but are difficult to shift. Small industrial loads, by contrast, offer a degree of flexibility and can be partially transferred or curtailed. These flexible loads can participate in demand DR, which is broadly divided into price-based (PBDR) and incentive-based (IBDR) programs [46,47]. In this context, only specific small industrial loads are eligible for PBDR, whereas EV charging can only be regulated through PBDR. Regarding response direction, residential loads can only provide load reduction, while small industrial and EV charging loads are capable of both load increase and decrease.\\
(1) Model of rural energy storage model\\

In the proposed system model. The generalized rural energy storage model is followed.
\begin{equation}
\begin{cases}
Q_t^{st}=Q_{t-1}^{st}+\left(\eta_{\mathrm{ch}}^{st}	P_{st,t}^{\mathrm{ch}}-\frac{P_{st,t}^{\mathrm{dis}}}{\eta_{\mathrm{dis}}^{st}}\right)\Delta t \\
Q_0^{st}=Q_T^{st} & 
\end{cases},st\in ST,t\in T\label{eq9}
\end{equation}
(2) Model of rural electric DR\\

 Therefore, the load DR model for rural users can be formulated as:
\begin{equation}
\Delta L^{\mathrm{e}}_{\mathrm{dr},t}=\sum_z\left(u_{\mathrm{PB,t}}^z\Delta L_{\mathrm{PB},t}^z+u_{\mathrm{IB},t}^z\Delta L_{\mathrm{IB},t}^z\right)
\label{eq23}
\end{equation}
PBDR primarily relies on time-of-use electricity pricing to guide users in adjusting controllable loads, thereby achieving peak shaving and valley filling [48]:\\

The detailed IBDR model is defined as:
\begin{equation}
\Delta L_{i,t}^{\mathrm{PB}} = L_{i,t}^0 \times \left( e_i^t \times \frac{\Delta P_i^t}{P_i^t} + \sum_{s=1,s\neq t}^{24} e_{i,s}^t \times \frac{\Delta P_i^s}{P_j^s} \right)
\label{eq24}
\end{equation}
In contrast, IBDR is directly triggered by the dispatching center, where users are contracted to provide response services in exchange for compensation. When power supply-demand imbalances occur, the operator activates user resources according to a step-by-step quotation curve, and users adjust their load upward or downward in accordance with the agreed response price. The IBDR model consists of three main types: contract response, increase response, and decrease response, and is formulated as:
\begin{equation}
\Delta L_{\mathrm{lB},t}^z=\Delta L_{\mathrm{lB},t}^{\mathrm{z,c}}+\sum_{m=1}^M\left(\Delta L_{\mathrm{lB},m,t}^{\mathrm{z,+}}+\Delta L_{\mathrm{lB},m,t}^{\mathrm{z,-}}\right)
\label{eq24}
\end{equation}
\noindent Where $\Delta L^{z}_{\mathrm{lB},t}$ represents the total transferable load (kWh); 
$\Delta L^{z,c}_{\mathrm{lB},t}$ denotes curtailed load (kWh); 
$\Delta L^{z,+}_{\mathrm{lB},m,t}$ and $\Delta L^{z,-}_{\mathrm{lB},m,t}$ are the load increase and reduction amounts (kWh).\\
(3) Model of rural electricity DR\\

Rural households often exhibit a considerable tolerance for moderate fluctuations in indoor temperature, suggesting a relatively low sensitivity to variations in heating or cooling demand. To quantitatively evaluate thermal comfort under these conditions, the Predicted Mean Vote (PMV) index is employed as a widely recognized metric [49]. The PMV is calculated as follows:
\begin{equation}
\lambda_{\mathrm{PMV}}=2.43-\frac{3.76(T_\mathrm{s}-T_{\mathrm{in}}^t)}{M(I_{\mathrm{cl}}+0.1)}
\label{eq23}
\end{equation}
The expected range of changes in the average evaluation indicators during the entire scheduling cycle is as follows:
\begin{equation}
\begin{cases}
\mid\lambda_{\mathrm{PMV}}\mid\leqslant0.9,t\in[1:00-7:00]\cup[20:00-24:00] \\
\lambda_{\mathrm{PMV}}\leqslant0.5,t\in[8:00-19:00] & 
\end{cases}
\label{eq23}
\end{equation}
The heat balance equation between heat load demand and temperature is:
\begin{equation}
\frac{\mathrm{d}T_{\mathrm{in}}^t}{\mathrm{d}t} = \frac{P_H^t - (T_{\mathrm{in}}^t - T_{\mathrm{out}}^t)KF}{\alpha_{\mathrm{air}}\rho_{\mathrm{air}}V}
\label{eq23}
\end{equation}

\noindent Where $\lambda_{\mathrm{PMV}}$ is the predicted mean vote index; 
$T_s$ and $T_{\mathrm{in}}^t$ denote surface and indoor temperatures (°C); 
$M$ is the metabolic rate; 
$I_{\mathrm{cl}}$ is the clothing insulation coefficient; 
$T^{t}_{\mathrm{out}}$ is outdoor temperature (°C); 
$K_F$, $\alpha_{\mathrm{air}}$, $\rho_{\mathrm{air}}$, and $V$ represent heat transfer coefficient, air convection coefficient, air density, and room volume.

It is established that integrated DR strategies can effectively coordinate cooling, heating, and electricity loads, underscoring their utility in multi-energy systems. In rural settings, thermal loads similarly possess significant spatiotemporal flexibility, with their DR potential manifesting mainly as load shifting and load reduction. By leveraging the coupling among different energy forms [93], the model formulated in this chapter integrates both shiftable and reducible thermal loads, as expressed by the following equations:
\begin{equation}
\left\{
\begin{array}
{l}{P_{s,h}^{\min,t}\leqslant\vert P_{s,h}^{\prime}\vert\leqslant P_{s,h}^{\max,t}} \\
{P_{h}^{t}=P_{p,h}^{t}+P_{s,h}^{t}-P_{c,h}^{t}} \\
{0\leqslant P_{c,h}^{t}\leqslant P_{c,h}^{\max,t}} \\
{\sum_{t=1}^{T}P_{s,h}^{t}=0}
\end{array}\right.
\label{eq23}
\end{equation}

\noindent Where $P_{s,h}^{\min,t}$ and $P^{\max,t}_{s,h}$ are the minimum and maximum transferable heat loads; 
$P^{t}_{p,h}$ and $P^{t}_{c,h}$ are the peak and curtailed thermal loads.

\subsection{Model of RIES system}\label{subsec2}
(1) Electricity purchasing from upstream grids\\

Calculate the carbon emissions generated by purchasing electricity from upstream power grids.
\begin{equation}
E_t^{\mathrm{buy}}=f^{\mathrm{grid}}p_t^{\mathrm{buy,e}}\label{eq8}
\end{equation}

\noindent Where $E^{\mathrm{buy}}_{t}$ denotes the carbon emission of purchased electricity (kgCO$_2$); 
$f_{\mathrm{grid}}$ is the emission factor of the upstream grid (kgCO$_2$/kWh); 
$p^{\mathrm{buy,e}}_{t}$ is the purchased power (kW).\\
(2) Power balance:\\

The operation of the proposed integrated system must maintain a real-time balance between energy supply and demand. The electric power balance equation is shown in Eq.20.\\
\begin{equation}
\hspace{-1.5cm}
\begin{aligned}
&P_{\mathrm{wt},t} + P_{\mathrm{pv},t} + P_{\mathrm{chp},t} + P_{\mathrm{gt},t} + P_{\mathrm{grid},t}
+ \big(P^{\mathrm{dis}}_{\mathrm{sto},t} - P^{\mathrm{ch}}_{\mathrm{sto},t}\big) \\
&\quad - P_{\mathrm{eb},t}^{\mathrm{e}} - P_{\mathrm{b2g},t} - P_{\mathrm{p2g},t}
= L^{0}_{\mathrm{e},t} + \Delta L^{\mathrm{e}}_{\mathrm{dr},t}
\end{aligned}
\label{eq13}
\end{equation}
\noindent Where $P_{\mathrm{wt},t}$, $P_{\mathrm{pv},t}$, $P_{\mathrm{chp},t}$, $P_{\mathrm{gt},t}$, $P_{\mathrm{grid},t}$, $P^{\mathrm{dis}}_{\mathrm{sto},t}$, and $P^{\mathrm{ch}}_{\mathrm{sto},t}$ represent generation and storage powers (kW); 
$ L^{0}_{\mathrm{e},t}$ and $ L^{0}_{\mathrm{h},t}$ are the base electric-thermal loads (kW); 
$\Delta L^{\mathrm{e}}_{\mathrm{dr},t}$ and $\Delta L^{\mathrm{h}}_{\mathrm{dr},t}$ are DR variations (kW); \\
(3) Thermal balance:\\
\begin{equation}
\begin{split}
Q_{\mathrm{gb},t}
+ Q_{\mathrm{eb},t}
+ Q_{\mathrm{chp},t}
+ Q_{\mathrm{whb},t} \\
+ \big(Q^{\mathrm{dis}}_{\mathrm{sto},t} - Q^{\mathrm{ch}}_{\mathrm{sto},t}\big)
= L^{0}_{\mathrm{h},t} + \Delta L^{\mathrm{h}}_{\mathrm{dr},t}
\end{split}
\label{eq:thermal_balance}
\end{equation}

\noindent Where $Q_{\mathrm{gb},t}$, $Q_{\mathrm{eb},t}$, $Q_{\mathrm{chp},t}$, and $Q_{\mathrm{whb},t}$ denote thermal outputs (kW).\\
(4) Gas balance:\\
\begin{equation}
F_{\mathrm{purchase},t}
+ F_{\mathrm{ng},t}
=
F_{\mathrm{gt},t}
+ F_{\mathrm{gb},t}
+ F_{\mathrm{chp},t}
\label{egas_balance}
\end{equation}
\noindent Where $F_{\mathrm{purchase},t}$ is the purchased natural gas (Nm$^3$); 
$F_{\mathrm{ng},t}$, $F_{\mathrm{gt},t}$, $F_{\mathrm{gb},t}$, and $F_{\mathrm{chp},t}$ are the gas consumption of each device (Nm$^3$).\\
(5) Carbon emission balance\\

Referring to the energy balance constraint, the carbon dioxide produced or absorbed by different parts of the proposed system should also be constrained. According to the IPCC and multiple national carbon-market regulations, the carbon dioxide emitted from biomass combustion is not counted as direct emissions from the energy system; instead, it is regarded as part of the biogenic carbon cycle and treated as net-zero emissions.
\begin{equation}
\begin{cases}
E_t^\mathrm{total}=E_t^\mathrm{out}+E_t^\mathrm{buy} \\
E^{\mathrm{GT}}_{t}+E_t^\mathrm{gb}-E_t^\mathrm{out}=0 & 
\end{cases}\label{eq13}
\end{equation}

The first line of formula (23) represents the carbon balance at the system level. \noindent Where $E_t^\mathrm{total}$ denotes the total system carbon emissions (kgCO$_2$); 
$E_t^\mathrm{out}$ is the carbon output; 
$E_t^\mathrm{buy}$, $E^{\mathrm{GT}}_{t}$, and $E_t^\mathrm{gb}$ are carbon emissions from purchased electricity, GT, and GB respectively.\\
(6) Other constraints\\

The RIES must also comply with some other constraints during operation, including capacity constraints, non-negativity constraints, state constraints, and upper/lower limits of unit ramping, as shown in Eq.24 \& 25.
\begin{equation}
\left\{
\begin{array}{l}
\underline{P_{\mathrm{gb}}}\leq P_{\mathrm{gb},t}\leq\overline{P_{\mathrm{gb}}} \\
0\leq P_{\mathrm{eb},t}\leq\overline{P_{\mathrm{eb}}} \\
0\leq p_{t}^{\mathrm{buy},\mathrm{e}} \\
0\leq V_{t}^{\mathrm{buy},\mathrm{g}} \\
\end{array}\right.
\label{eq20}
\end{equation}
\begin{equation}
\begin{cases}
\underline{R_{\mathrm{gb}}}\leq P_{\mathrm{gb},t}-P_{\mathrm{gb},t-1}\leq\overline{R_{\mathrm{gb}}} \\
\underline{R_{\mathrm{chp}}}\leq P_{\mathrm{chp},t}-P_{\mathrm{chp},t-1}\leq\overline{R_{\mathrm{chp}}} \\
\end{cases}
\label{eq21}
\end{equation}
\noindent Where $P_{\mathrm{gb},t}$, $P_{\mathrm{eb},t}$, and $p_{t}^{\mathrm{buy},\mathrm{e}}$ are the power or heat outputs of devices (kW); 
$V_{t}^{\mathrm{buy},\mathrm{g}}$ is the purchased gas volume (Nm$^3$); 
$R_{\mathrm{gb}}$ and $R_{\mathrm{chp}}$ are the ramp rates of boiler and CHP units (kW/h).\\
(7) Renewable and biomass energy constraints\\

Renewable energy sources, including WP, PV generation, and biomass-derived energy, are integrated into the system with different characteristics. Wind and solar outputs exhibit inherent variability due to weather conditions and require a chance-constrained approach to ensure reliable operation.\\
\begin{equation}
0 \le P_{i,t}^{\mathrm{RES}} \le P_{i,t,\mathrm{fore}}^{\mathrm{RES}} + \sigma_{i,t,\mathrm{fore}} \phi_a^{-1}(1-\eta)
\label{eq:RES_uncertainty}
\end{equation}

Biomass energy supply is relatively controllable, and its utilization is constrained only by the installed capacity limits of the CHP units. This constraint ensures reliable operation while fully utilizing the biomass-derived syngas for power and heat generation.
\begin{equation}
0 \le G_{\mathrm{chp},t} \le G_{\mathrm{chp},\mathrm{max}}
\label{eq:CHP_biomass_only}
\end{equation}

\noindent Where $P_{i,t}^{\mathrm{RES}}$ is the renewable output (kW); 
$P^{\mathrm{RES}}_{i,t,\mathrm{fore}}$ is the forecasted value (kW); 
$\sigma_{i,t,\mathrm{fore}}$ is the forecasting deviation; 
$\phi^{-1}_{a}(1-\eta)$ denotes the inverse standard normal distribution for confidence level $(1-\eta)$; 
$G_{\mathrm{chp},t}$ and $G_{\mathrm{chp},\mathrm{max}}$ are the real and maximum gas supply for CHP.

\section{Optimization strategy for carbon reduction in RIES}\label{sec3}
\subsection{Strategy 1: Carbon reduction strategy considering carbon trading mechanism}\label{subsec2}
\noindent\hspace{1em}
This section examines the operation of a RIES under a carbon trading mechanism. The mathematical modeling of this trading mechanism typically consists of two parts: a carbon quota allocation model and a tiered carbon trading model, which together describe the constraints and strategic choices in the quota allocation and market game process.
\subsubsection{Carbon emission quota model}\label{subsubsec2}
\noindent\hspace{1em}
At present, in China’s carbon quota management practice, the most common allocation method is the free allocation mechanism, which means that initial carbon emission quotas can be obtained without paying any fees. The quota model is as follows:
\begin{equation}
\begin{cases}
E_{\mathrm{e,buy}} = \lambda_{\mathrm{e}} \sum_{t=1}^{T} P_{\mathrm{e,buy}}(t) \\
E_{\mathrm{GB}} = \lambda_{\mathrm{g}} \sum_{t=1}^{T} P_{\mathrm{e,GB}}(t) \\
E_{\mathrm{IES}} =E_{\mathrm{e,buy}} + E_{\mathrm{GB}}
\end{cases}
\label{eq:energy_system}
\end{equation}

\noindent Where $E_{\mathrm{e,buy}}$ and $E_{\mathrm{GB}}$ denote the carbon quota for purchased electricity and gas boiler (kgCO$_2$); 
$\lambda_{\mathrm{e}}$ and $\lambda_{\mathrm{g}}$ are their carbon intensity coefficients; 
$P_{\mathrm{e,buy}}(t)$ and $P_{\mathrm{e,GB}}(t)$ are energy consumptions (kWh); 
$E_{\mathrm{IES}}$ is the total carbon allowance for the IES.

\subsubsection{Tiered carbon emissions trading model}\label{subsubsec2}
\noindent\hspace{1em}Once the carbon emission quota for the RIES is determined, its actual trading volume in the carbon market is subsequently determined. To actively promote carbon reduction and guide the system operator towards low-carbon strategies, this paper introduces a tiered carbon trading pricing mechanism [50,51]. This model categorizes the demand for carbon allowances into several tiers, with the unit price escalating as the purchase volume rises. This escalating cost structure imposes a market-driven constraint that strengthens with increased emission demand. The cost model is formulated as follows:
\begin{equation}
C^{\mathrm{CET}}=
\begin{cases}
\beta E_\mathrm{IES}, & E_\mathrm{IES}\leq l \\
\beta(1+\zeta)(E_\mathrm{IES}-l)+\beta l, & l < E_\mathrm{IES}\leq 2l \\ 
\beta(1+2\zeta)(E_\mathrm{IES}-2l)+\beta(2+\zeta)l, & 2l < E_\mathrm{IES}\leq 3l \\ 
\beta(1+3\zeta)(E_\mathrm{IES}-3l)+\beta(3+3\zeta)l, & 3l < E_\mathrm{IES}\leq 4l \\ 
\beta(1+4\zeta)(E_\mathrm{IES}-4l)+\beta(4+6\zeta)l, & E_\mathrm{IES} > 4l 
\end{cases}
\label{eq23}
\end{equation}

\noindent Where $C^{\mathrm{CET}}$ denotes the total carbon trading cost (¥); 
$\beta$ is the base carbon price; 
$\zeta$ is the incremental coefficient; 
$l$ is the emission quota limit; 
$E_{\mathrm{IES}}$ is the total system emissions (kgCO$_2$).
\subsection{Strategy 2: Carbon reduction strategy considering electricity and heat DR}\label{subsec2}
\noindent\hspace{1em}Integrated DR has been demonstrated as an effective approach for coordinating cooling, heating, and electricity loads in multi-energy systems, highlighting its significant practical value. Building upon this foundation, this study notes that thermal loads in rural contexts also exhibit considerable spatiotemporal flexibility, similar to electricity loads. This flexibility translates into two primary forms of DR potential: load shifting and load reduction. Leveraging the coupling characteristics of multiple energy carriers, the thermal DR model developed in this chapter incorporates both shiftable and reducible loads, formulated as follows:
\begin{equation}
\left\{
\begin{array}{l} 
{P_{\mathrm{s},\mathrm{h}}^{\mathrm{min},t}\leqslant\vert P_{\mathrm{s},\mathrm{h}}^{\prime}\vert\leqslant P_{\mathrm{s},\mathrm{h}}^{\mathrm{max},t}} \\
{P_{\mathrm{h}}^{t}=P_{\mathrm{p},\mathrm{h}}^{t}+P_{\mathrm{s},\mathrm{h}}^{t}-P_{\mathrm{c},\mathrm{h}}^{t}} \\
{0\leqslant P_{\mathrm{c},\mathrm{h}}^{t}\leqslant P_{\mathrm{c},\mathrm{h}}^{\mathrm{max},t}} \\
{\sum_{t=1}^{T}P_{\mathrm{s},\mathrm{h}}^{t}=0}
\end{array}\right.
\label{eq23}
\end{equation}
\noindent Where $C_{\chi}$ is the total system operation cost (¥); 
$C^{\mathrm{CET}}$, $C_{\mathrm{buy}}$, and $C_{\mathrm{O}}$ are the carbon trading, energy purchase, and operation maintenance costs (¥).

\subsection{Strategy 3: Carbon reduction strategy considering carbon trading and DR}\label{subsec2}
\noindent\hspace{1em}Guided by the principle of source-load coordination, the research framework of this subsection is illustrated in Fig. 3. The model aims to achieve low-carbon operation of the IES through coordinated dynamic electricity pricing and integrated electro-thermal DR. Within the established bi-level optimization framework, the upper level represents the main grid, which optimizes dynamic electricity prices via an economic dispatch model to minimize total operating costs under a system-wide carbon emission cap. The lower level, represented by the IES, responds to price and low-carbon signals from the grid by participating in multiple energy markets. Its objective is to maximize consumer surplus while accommodating users' preferences for low-carbon energy. Through dynamic interaction and game-theoretic processes between the two levels, efficient resource coordination and synergy among source, grid, and load are achieved, ultimately enhancing the low-carbon economic performance of the RIES.
\begin{figure}[h]
\centering
\includegraphics[width=0.8\textwidth]{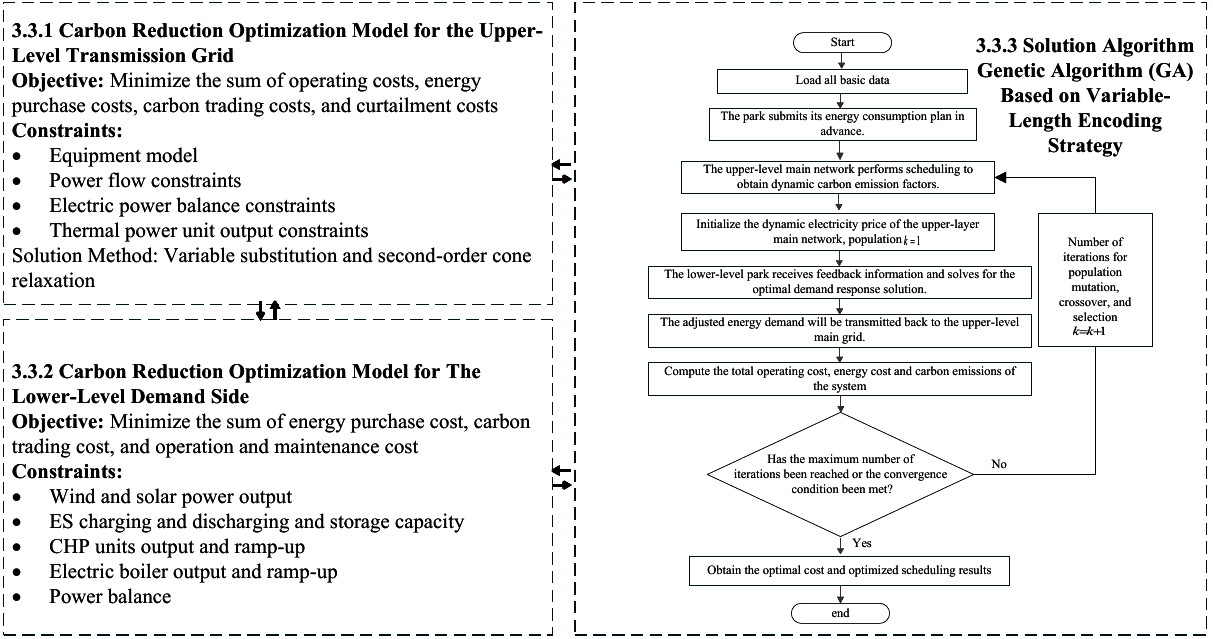}
\caption{Framework of the carbon reduction optimization model and solution algorithm for RIES}\label{fig1}
\end{figure}

\subsubsection{Carbon reduction optimization model for the upper-level transmission grid}\label{subsubsec2}
(1) Main grid carbon reduction optimization target\\

The main grid carbon reduction optimization target expression is as follows:
\begin{equation}
\begin{cases}
\min C_{\mathrm{main}}=C_{\mathrm{coal}}+C_{\mathrm{gas}}+C_{\mathrm{cur}}+C_{\mathrm{market}}-C_{\mathrm{sell}} \\
C_{\mathrm{coal}}=\sum_{n_{\mathrm{h}}=1}^{N_{\mathrm{h}}}\sum_{t=1}^{T}\phi_{\mathrm{CF},i,t}\eta_{\mathrm{CF},i,t}(a_{i}P_{\mathrm{CF},t}^{2}+b_{i}P_{\mathrm{CF},t}+c_{i}) \\
C_{\mathrm{gas}}=\sum_{n_{\mathrm{g}}=1}^{N_{\mathrm{g}}}\sum_{t=1}^{T}\phi_{\mathrm{GA},i,t}\eta_{\mathrm{GA},i,t}q_{i}P_{\mathrm{GA},t} \\
C_{\mathrm{cur}}=\sum_{t=1}^{T}\phi_{\mathrm{cur}}(P_{\mathrm{PV},\mathrm{Z},\mathrm{max},t}-P_{\mathrm{PV},Z,t}) \\
C_{\mathrm{sell}}=\sum_{i=1}^{N_{y}}(\sum_{t=1}^{T}(\phi_{e,t}P_{\mathrm{buy},i,t})+\phi_{\mathrm{cur}}\sum_{t=1}^{T}Q_{\mathrm{buy},i,t})
\end{cases}
\label{eq23}
\end{equation}
(2) Main grid carbon reduction constraints\\

The main grid constraints are explained as follows:
\begin{equation}
\left\{
\begin{aligned}
&\sum_{ik \in \Omega_{\text{branch}}} P_{t,ik} - \sum_{ji \in \Omega_{\text{branch}}} (P_{t,ji} - r_{ji}I_{t,ji}^2) = P_{t,i} \\
&\sum_{ik \in \Omega_{\text{branch}}} Q_{t,ik} - \sum_{ji \in \Omega_{\text{branch}}} (Q_{t,ji} - x_{ji}I_{t,ji}^2) = Q_{t,i} \\
&U_{t,i}^2 = U_{t,j}^2 + 2(r_{ij}P_{t,ij} + x_{ij}Q_{t,ij}) - (r_{ij}^2 + x_{ij}^2)I_{t,ij}^2 \\
&I_{t,ij}^2U_{t,i}^2 = P_{t,ij}^2 + Q_{t,ij}^2 \\
&P_{t,i} = P_{t,i}^{\text{WT}} + P_{t,i}^{\text{PV}} + P_{t,i}^{\text{DG}} - P_{t,i}^{\text{L}} + P_{t,i}^{\text{ESS}} \\
&Q_{t,i} = Q_{t,i}^{\text{DG}} - Q_{t,i}^{\text{L}}
\end{aligned}
\right.
\label{eq23}
\end{equation}

\subsubsection{Carbon reduction optimization model for the lower-level carbon market and DR}\label{subsubsec2}

(1) Demand side optimization objective\\

The optimization objective is to minimize the sum of the regional energy purchase cost , carbon trading cost, and operation and maintenance cost, which can be expressed as:
\begin{equation}
\min C_{\chi}=C^{\mathrm{CET}}+C^{\mathrm{buy}}+C^{\mathrm{O}}
\end{equation}
(2) Constraints\\

The remaining equipment constraints in the IES are presented in Section 2.2.
\subsubsection{Model solving algorithm}\label{subsubsec2}
\noindent\hspace{1em}For the nonlinear two-level programming model established in this paper, since the main grid needs to formulate an adaptive electricity selling price in the upper optimization model of the model, it constitutes a complex nonlinear optimization problem. This chapter uses a genetic algorithm based on a variable-length coding strategy as an optimization tool to effectively improve the diversity of solutions and global optimization capabilities [52,53]. In the lower optimization model, the IES interacts with the main grid as a decision-making body and uses a distributed collaborative scheduling method to solve the optimization problem. The relevant solution process and implementation steps are detailed description below. \\
\begin{algorithm}[h]
\caption{Solution Algorithm for the Two-Level RIES Optimization Model}\label{algo2}
\begin{algorithmic}[1]
\Require Upper-level main grid and lower-level park parameters
\Ensure Optimal electricity selling price and park energy consumption plan
\State Initialize GA parameters: population size, max generations, mutation rate, etc.
\State Generate initial population using variable-length encoding for electricity prices
\While{convergence condition not met \textbf{and} generation $< $ max generations}
    \For{each individual in population}
        \State Evaluate fitness of upper main grid (total operating cost including carbon trading)
        \State Optimize lower park energy consumption plan based on electricity price and carbon intensity
        \State Feedback results to upper main grid
        \State Update source-side output and calculate total cost
    \EndFor
    \State Select high-fitness individuals as parents
    \State Apply mutation and recombination to generate new population
    \State $k \Leftarrow k + 1$
\EndWhile
\State Output optimal electricity selling price, main grid total cost, and park energy consumption plan
\end{algorithmic}
\end{algorithm}
(1) Initialization\\

Initialize the main grid operating parameters, including the population size, maximum number of generations, mutation rate and other parameters of the genetic algorithm. The upper main grid uses a genetic algorithm based on a variable-length coding strategy to randomly generate the initial electricity selling price of the group main grid, and performs preliminary scheduling based on the initial energy consumption plan of the park, generates the time-varying average carbon emission intensity of the main grid, and transmits it to the lower park, and sends the number of generations to initialize.\\
(2) Fitness function design\\

The fitness function of the upper main grid includes the operating cost of coal-fired units, the operating cost of gas-fired units, the carbon trading cost of the main grid, the cost of curtailed solar power, and the revenue from energy sales. The goal is to minimize the total operating cost of the main grid. The fitness function of the lower park is defined as the sum of the energy purchase cost, carbon trading cost, and operation and maintenance cost. The goal is to minimize the total operating cost of park i.\\
(3) Population generation and encoding\\

The variable-length encoding strategy is used to represent the main grid's electricity sales price as a variable-length chromosome code. The chromosome length of each individual represents the dimension of the electricity sales price, and the gene value represents the specific value of the electricity sales price.\\
(4) Fitness evaluation\\

Evaluate the fitness of each individual: the upper main grid calculates the total cost of the main grid based on the individual electricity sales price. The lower park optimizes the energy consumption plan based on the main grid's electricity sales price and carbon emission intensity, and feeds the optimization results back to the upper main grid. The main grid optimizes the output of the source-side units based on the park's energy consumption plan, calculates the total cost, and retains the current cost and optimal incentive price. \\
(5) Selection and mutation\\

Selection operation: According to the value of the fitness function, select individuals with higher fitness from the current population as parents.\\
Mutation operation: Perform mutation operations on the parent individuals, including gene mutation (randomly changing the values of certain genes of the electricity selling price) and gene recombination (randomly exchanging gene fragments of the parent individuals). Generate a new incentive price, and the number of generations is $k = k + 1$\\
(6) \textrm{Iterative optimization}\\

Repeat steps (4) to (5) until the convergence condition is met (for example, the change in the fitness function is less than the preset threshold) or the maximum number of iterations is reached. In each iteration, update the population and record the optimal solution.\\
(7) Result output\\

After the convergence condition is met or the maximum number of generations is reached, output the optimization results, including the optimal electricity selling price, the total cost of the main grid, the energy consumption plan of RIES, etc.

\section{Case Study}\label{sec2}
\subsection{Data foundation}\label{subsec2}
\noindent\hspace{1em}
To validate the carbon reduction strategy for a RIES that considers carbon markets and DR, as proposed in this section, this section focuses on a RIES park located in northern China, using a 24-hour typical day as the operation cycle. The Biomass CHP plant, GB, and low-temperature waste HP generation device, EB, and HP are connected to the system. The parameters are shown in Table 1. The initial electrical and thermal loads of the system, as well as the predicted WP output, are shown in Fig.4. The natural gas cost is 3.45 yuan/m³, which translates to a unit calorific value price of 0.35 yuan/kWh. The time-of-use electricity price, as published by the Beijing Municipal Development and Reform Commission, is shown in Table 2. The base price for carbon trading is 0.3 yuan/kg, the interval length is 2000 kg, and the price increase rate is 25\%. The park purchases any electricity shortfall from the upstream power grid, and the natural gas required for the CHP unit is purchased from the upstream gas grid. The system operates on a heat-based electricity pricing model. There are many users in the park, and the overall energy consumption is high. The DR speed is fast and the response capability is strong, but the enthusiasm for participating in the electricity DR is low. The proportions of fixed, transferable, reducible and replaceable loads are set at 40\%, 35\%, 20\% and 5\% respectively; at the same time, a thermal load is set in RIES for DR.
\begin{table}[h]
\caption{Parameters of devices in RIES}\label{tab4-1}
\begin{tabularx}{\textwidth}{@{}p{0.19\textwidth}X@{}}
\toprule
Device & Main Parameters and Technical Characteristics \\
\midrule
Biomass CHP & Installed capacity 4000~kW; electrical efficiency 30\%; thermal efficiency 50\%\\
PG & Operating temperature 700~$^\circ$C; pyrolysis efficiency 82\%\\
B2G & Upgrading efficiency 90\%; purified biogas CH$_4$ content 96\%\\
Anaerobic digester & Conversion efficiency 70\%; optimal temperature 35~$^\circ$C\\
GT & Installed capacity 4000~kW; electrical efficiency 29\%; thermal efficiency 42\%\\
GB & Installed capacity 1000~kW; thermal efficiency 88\%\\
EB & Installed capacity 400~kW; efficiency 95\%\\
HP & Installed capacity 400~kW; coefficient of performance 300\%\\
ES & Electrical charge/discharge efficiency 85--90\%; thermal storage efficiency 85\%\\
\bottomrule
\end{tabularx}
\end{table}

\begin{table}[h]
\caption{System Time-of-Use Electricity Price}\label{tab4-2}
\begin{tabularx}{\textwidth}{@{}p{0.2\textwidth} X p{0.2\textwidth}@{}}
\toprule
Type & Time Interval & Price (yuan/kWh) \\
\midrule
Valley period & 23:00-07:00 & 0.2988 \\
Flat period & 07:00-10:00, 15:00-18:00, 21:00-23:00 & 0.5855 \\
Peak period & 10:00-15:00, 18:00-21:00 & 0.8882 \\
\bottomrule
\end{tabularx}
\end{table}

\begin{figure}[h]
\centering
\includegraphics[width=0.7\textwidth]{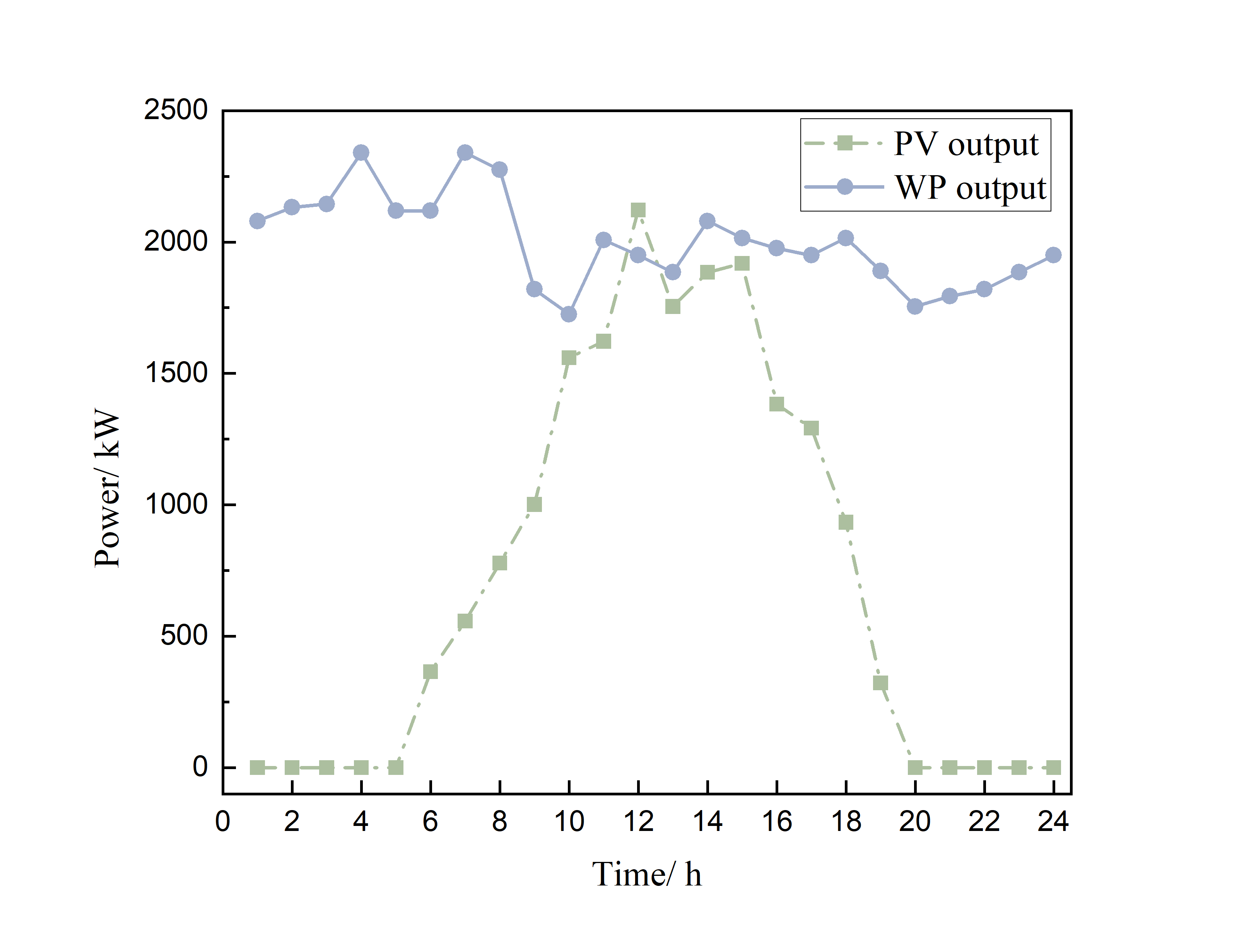}
\caption{Wind and Solar Power Output Curve}\label{fig1}
\end{figure}

\subsection{Scenario division}\label{subsec2}
To verify the feasibility of the RIES decarbonization strategy proposed in this chapter, which incorporates carbon markets and dynamic DR, this section presents four comparative analyses:\\
Scenario 1 (Baseline): Neither carbon markets nor dynamic DR are considered;\\
Scenario 2: Only electricity and thermal DR is considered;\\
Scenario 3: Only carbon trading mechanism is considered;\\
Scenario 4: Both carbon markets and electricity and thermal DR are considered.\\
The costs and actual carbon emissions of each scenario are shown in Table 3.

\begin{table}[h]
\caption{Comparison of Operation Results under Different Scenarios}\label{tab4-3}
\scriptsize
\begin{tabular}{@{}p{0.9cm} p{2.1cm} p{2.0cm} p{1.5cm} p{2.0cm} p{2.2cm}@{}}
\toprule
Scenario & Total Operation cost (¥) & Energy Purchase cost (¥) & Operation and Maintenance cost (¥) & Carbon trading cost (¥) & Carbon emissions (kg CO$_2$) \\
\midrule
1 & 56413.86 & 47131.34 & 3265.29 & 6017.23 & 62708.81 \\
2 & 55946.32 & 46060.44 & 3562.73 & 6323.15 & 61568.80 \\
3 & 51107.66 & 42372.80 & 3516.14 & 5218.72 & 58701.20 \\
4 & 49050.63 & 40405.72 & 3559.90 & 5085.01 & 56805.34 \\
\bottomrule
\end{tabular}
\end{table}
\vspace{-2em} 
\subsection{Analysis of carbon reduction potential results}\label{subsec2}
(1) Analysis of the carbon reduction potential of DR load regulation\\

Compared with scenario 1, scenario 2 introduces a time-of-use electricity price mechanism to guide users to adjust their electricity consumption behavior according to the electricity price period, thereby effectively achieving “peak shaving and valley filling", reducing the power load demand during peak periods, and alleviating the operating pressure of the source-side units. During periods with higher electricity prices, users actively reduce electricity consumption, effectively avoiding high-cost electricity consumption. At the same time, they increase load during periods with lower electricity prices or sufficient renewable energy output, thereby improving the utilization rate of clean energy such as WP and PV power, thereby reducing dependence on electricity purchases from the upper power grid and reducing the overall operating cost of the system. The simulation results show that the energy purchase cost of scenario 2 is 2.27\% lower than that of scenario 1. This cost advantage is mainly attributed to the introduction of a dynamic DR strategy, which enables the load to better match the energy supply structure, thereby selecting a more economical power source combination. \\

\begin{figure}[h]
\raggedleft
\includegraphics[width=0.9\textwidth]{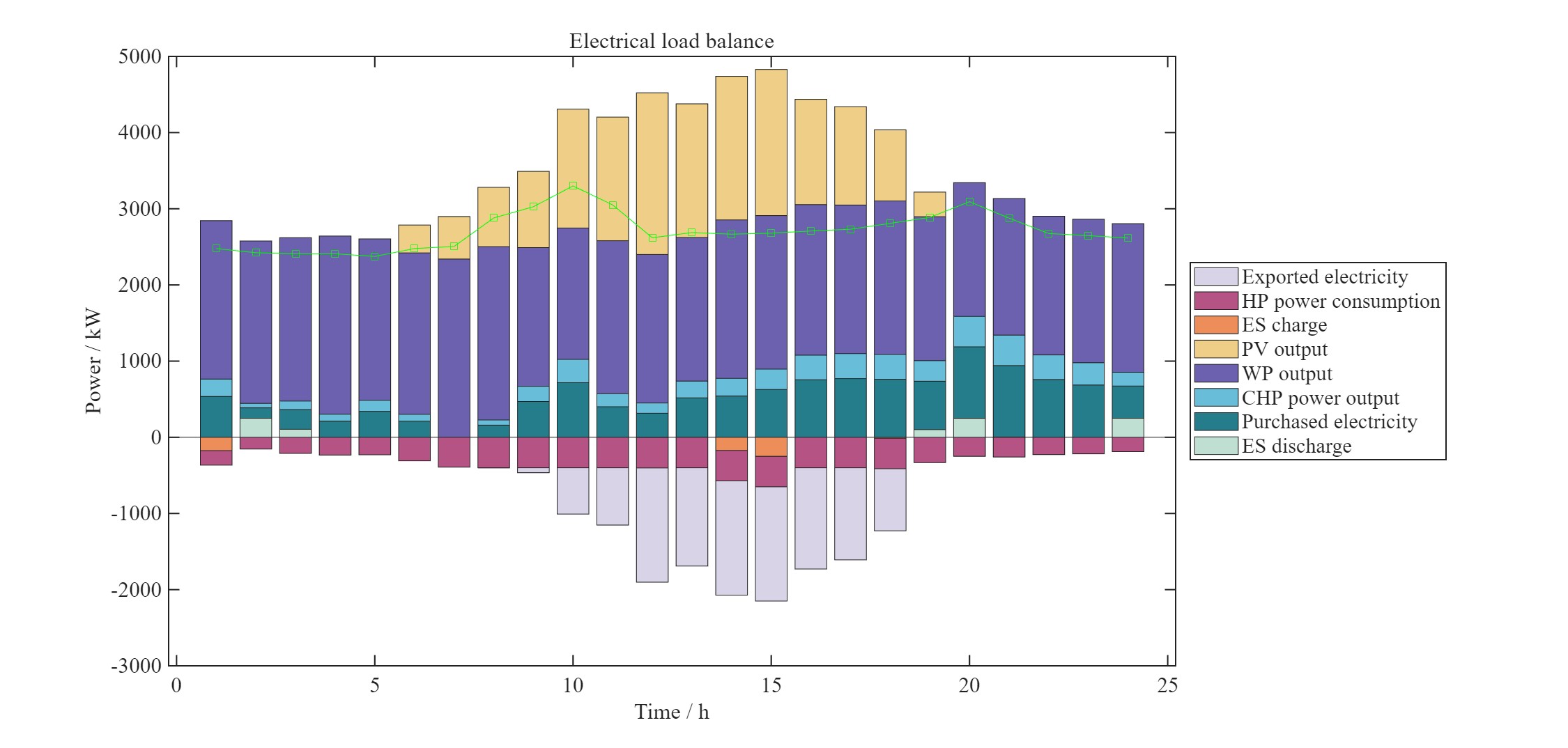}
\caption{Electric Load Balance Diagram of Scenario 1}\label{fig1}
\end{figure}

As shown in Fig.4 to 6, in terms of the specific load supply structure, electricity is mainly shared by WP, PV power and other renewable energy sources and GT, and grid power purchase only plays a supporting role. During the high-price period of 10:00 AM to 3:00 PM, due to abundant wind and solar resources, the system increased its share of green electricity, effectively curbing carbon emissions. Furthermore, between 4:00 PM and 12:00 PM, the CHP units released power in a timely manner, effectively alleviating power supply pressure during the evening peak demand period. Throughout the entire dispatch cycle, GT not only compensated for periods of insufficient wind and solar resources but also reduced the system's carbon emissions intensity by providing clean gas energy, mitigating the impact of peak loads on the grid. The final results showed that total carbon emissions in Scenario 2 were reduced by 1,140.01~\text{kg CO$_2$} compared to Scenario 1, demonstrating excellent low-carbon operation results.\\
\begin{figure}[h]
\raggedleft
\includegraphics[width=0.9\textwidth]{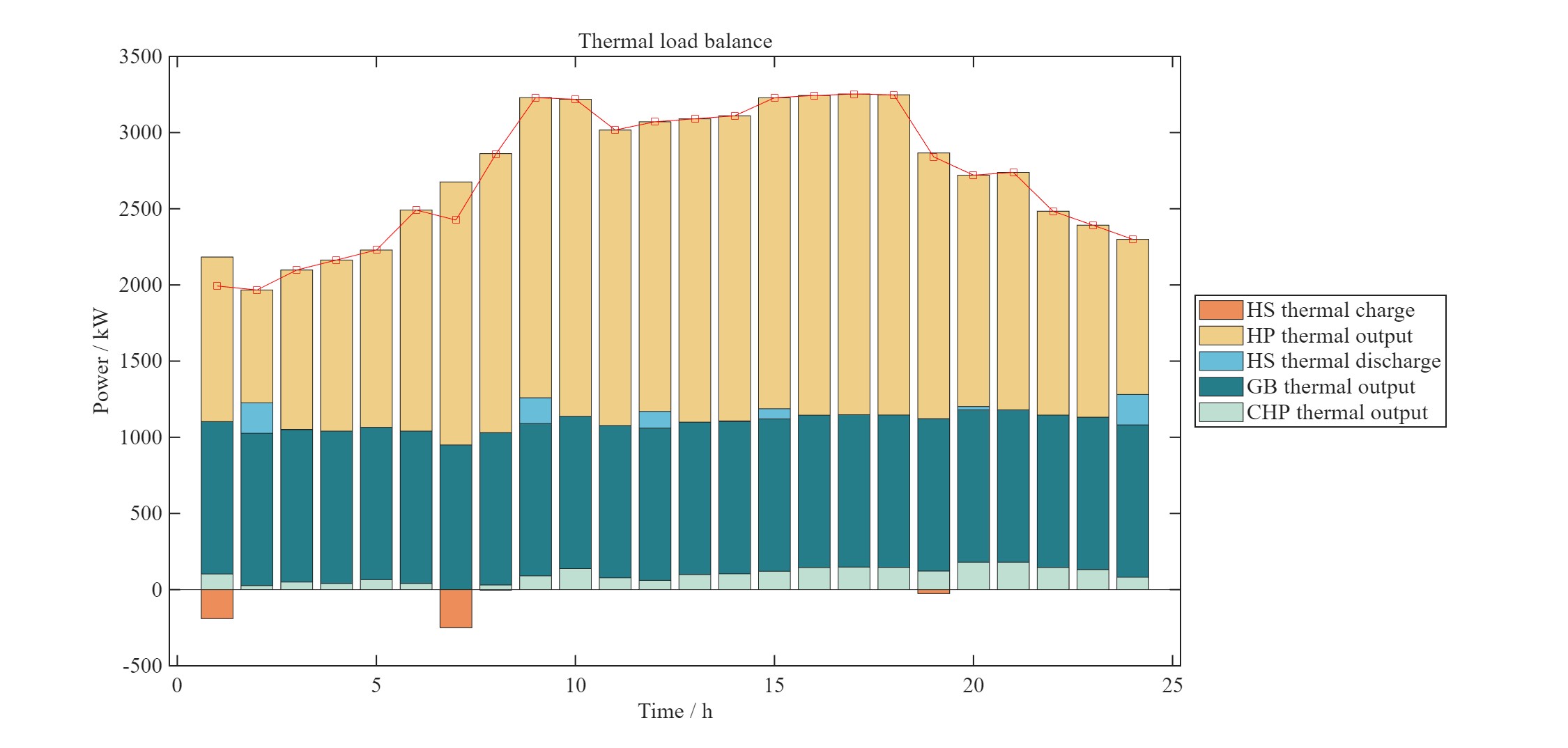}
\caption{Thermal Load Balance Diagram of Scenario 1}\label{fig1}
\end{figure}

As shown in Fig.7 and 9, a comparison of the load curves before and after optimization for Scenario 2 reveals that flexible loads with time-adjustable characteristics (including shiftable and curtailable loads) generally exhibit a trend of shifting from peak hours to flat and off-peak hours. This load shift not only helps improve the local absorption capacity of renewable energy but also effectively enhances the economic efficiency of system scheduling. Furthermore, the system selectively curtails load during periods of high electricity prices, primarily during peak demand periods, effectively alleviating system operating pressures. From an energy scheduling perspective, to minimize reliance on grid power purchases, the system prioritizes the use of wind and photovoltaic resources, maintaining nearly full power generation capacity for wind and solar power generation equipment.\\

\begin{figure}[h]
\centering
\includegraphics[width=0.64\textwidth]{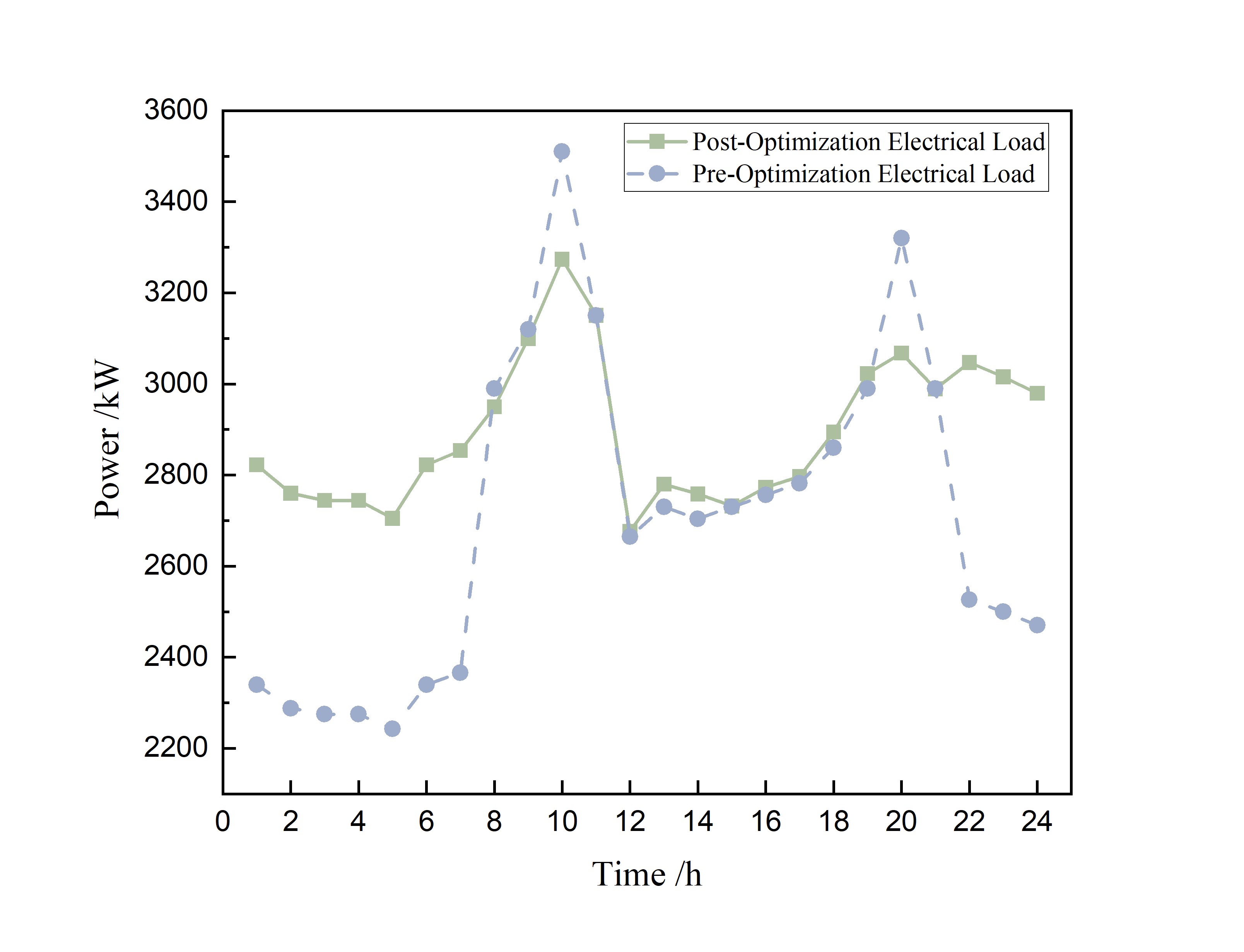}
\caption{Electric load curves before and after optimization in Scenario 2}\label{fig1}
\end{figure}

\begin{figure}[h]
\centering
\includegraphics[width=0.64\textwidth]{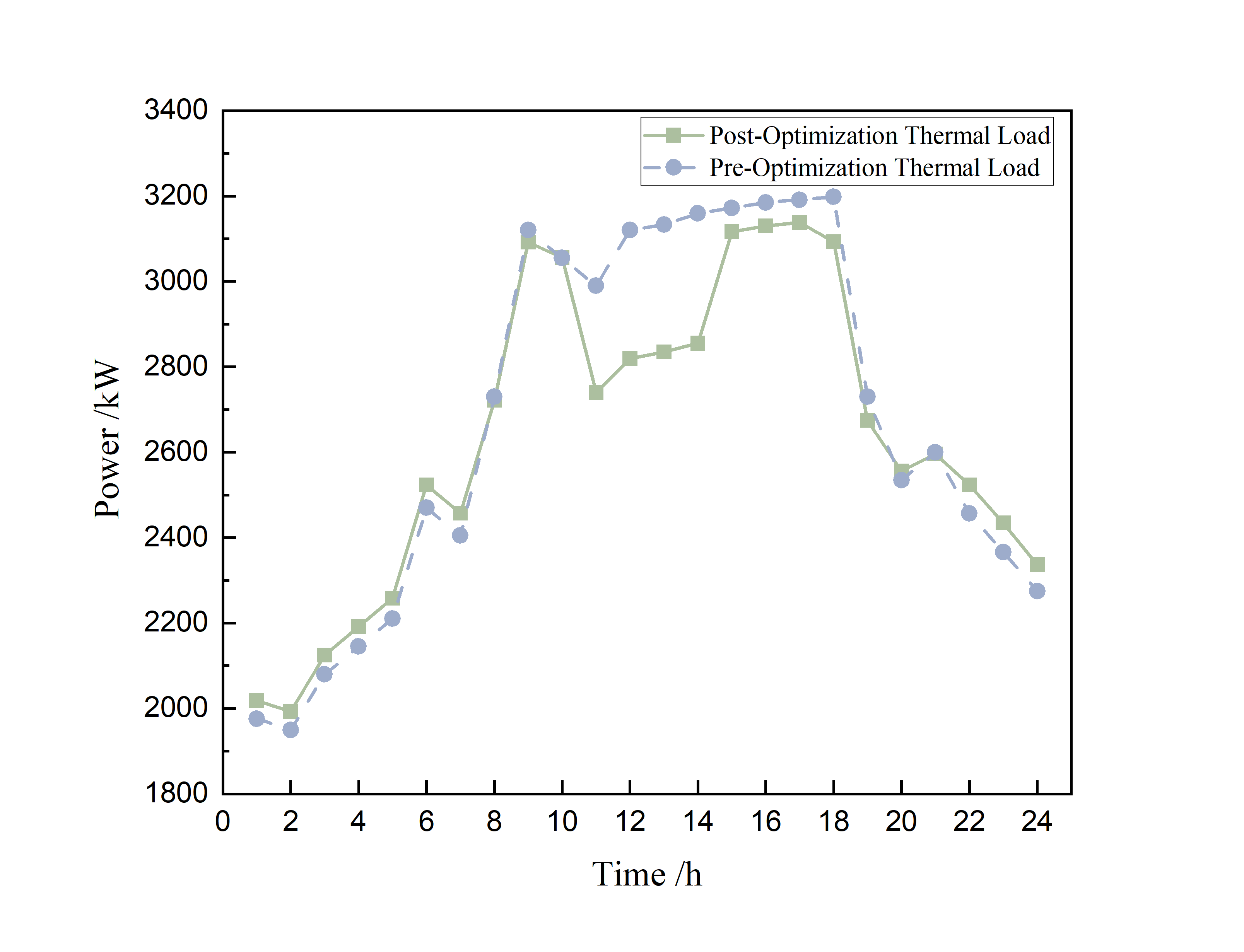}
\caption{Thermal load curves before and after optimization in Scenario 2}\label{fig1}
\end{figure}

As shown in Fig.8 and 10, during periods of low heat load, such as 12:00 PM to 9:00 AM and 10:00 PM to 12:00 AM, HP are used to convert and store heat energy in the thermal energy storage system. Given the abundant wind and solar resources and low electricity prices during these periods, this strategy not only fully utilizes clean energy but also provides a backup heat source for subsequent peak heat load periods. During periods of increased heat demand, the system prioritizes the use of stored heat energy to meet heating needs, effectively reducing reliance on direct operation of GB and HP, achieving the dual goals of energy conservation and emission reduction while lowering operating costs. Comparing the heat load curves before and after optimization further validates the effectiveness of this strategy. Similar to the electrical load, the movable heat load also exhibits a trend of shifting from peak to off-peak periods. This significantly alleviates the system's heating pressure during the evening peak heat load period while ensuring user thermal comfort. During off-peak hours, run the HP to store heat in the thermal energy storage system. During normal hours, if the heat load increases slightly, briefly run the HP to meet demand, avoiding frequent startup of GB. During off-peak hours, run HP or electric heating device to store heat in the thermal ES, reducing the proportion of GB usage. During peak hours, prioritize releasing stored heat to meet the base heat load. If the stored energy is insufficient, briefly run HP.\\

\begin{figure}[h]
\raggedleft
\includegraphics[width=0.9\textwidth]{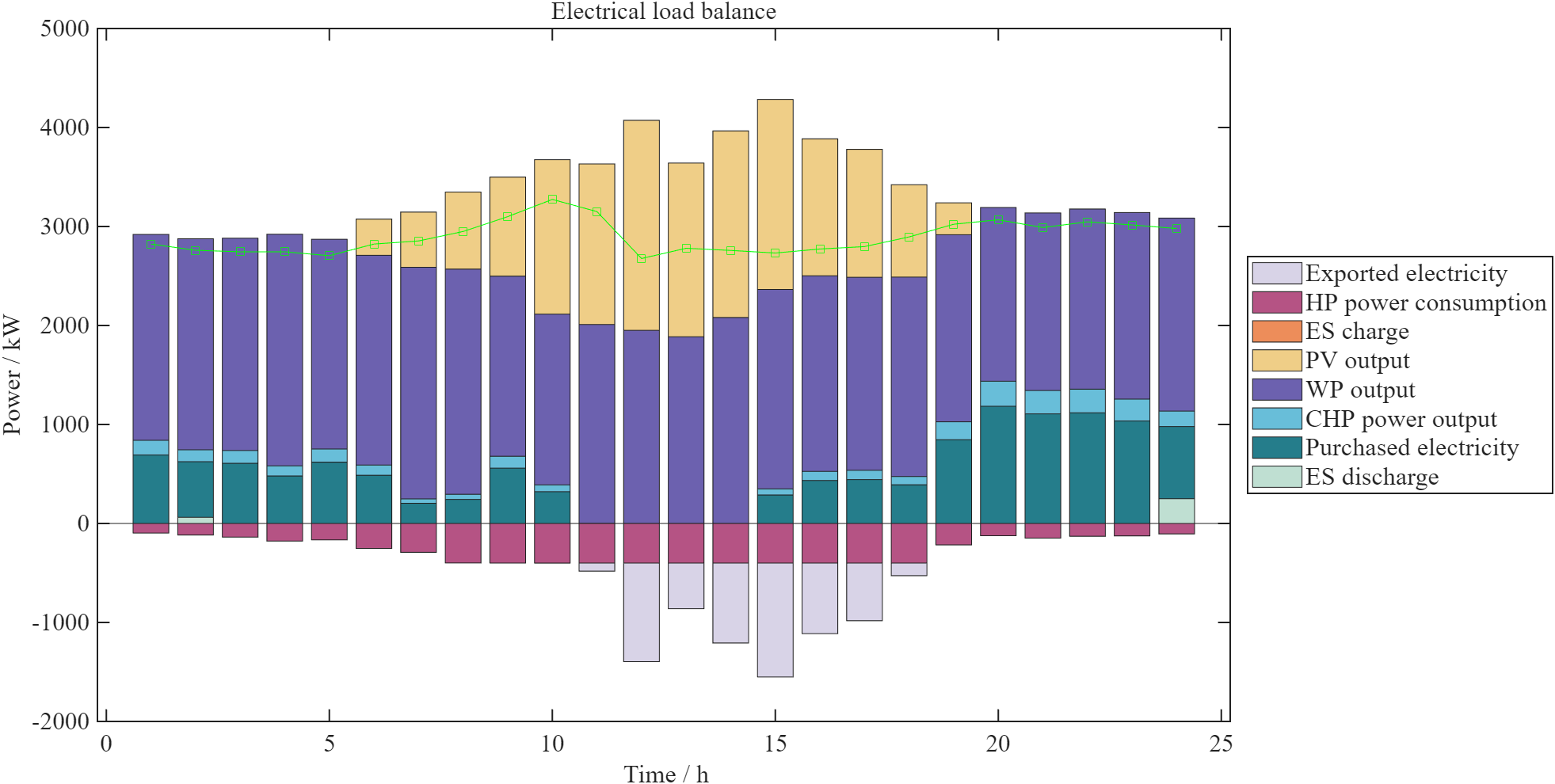}
\caption{Electric Load Balance Diagram of Scenario 2}\label{fig1}
\end{figure}

\begin{figure}[h]
\raggedleft
\includegraphics[width=0.9\textwidth]{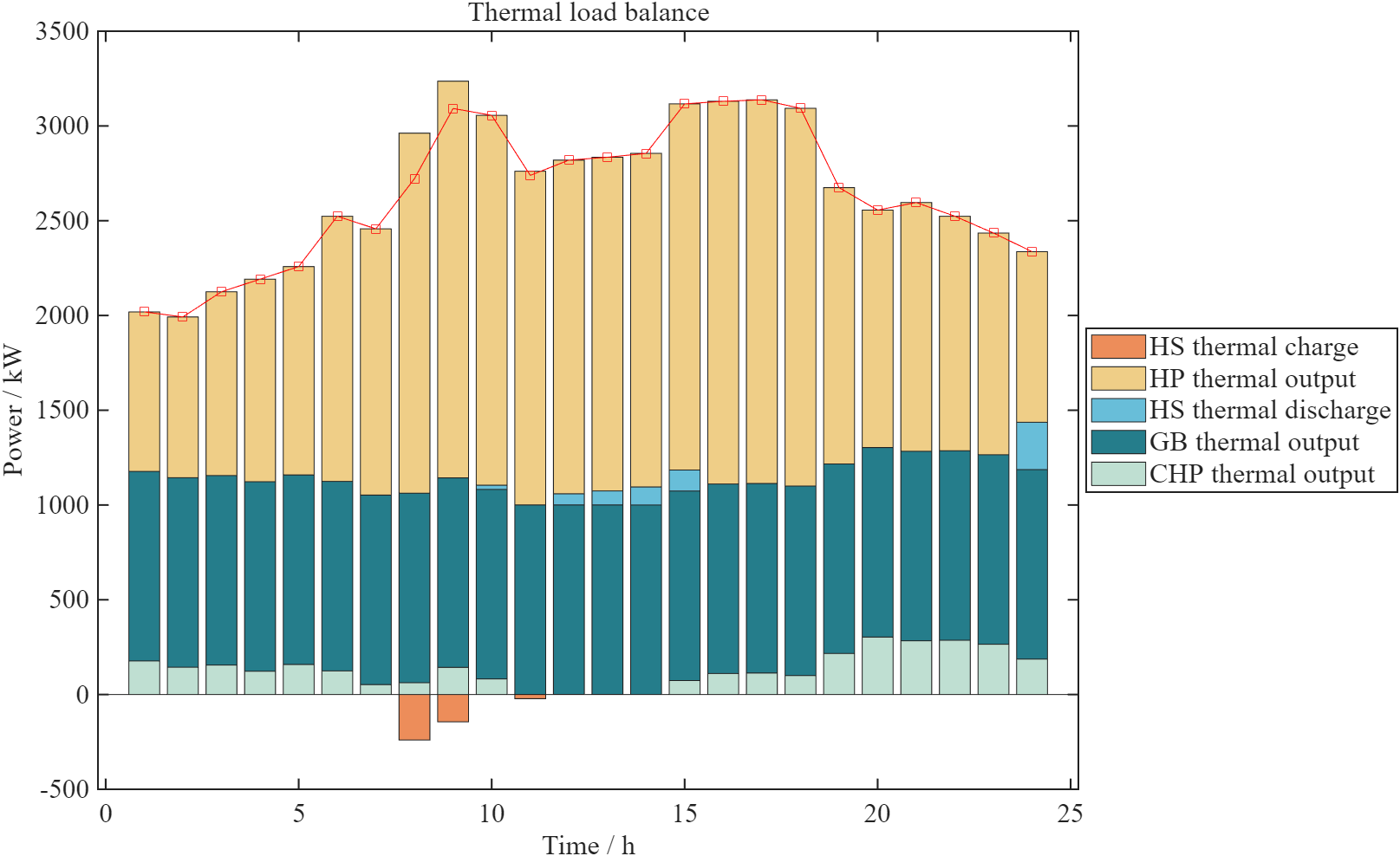}
\caption{Thermal Load Balance Diagram of Scenario 2}\label{fig1}
\end{figure}

\begin{figure}[h]
\centering
\includegraphics[width=0.64\textwidth]{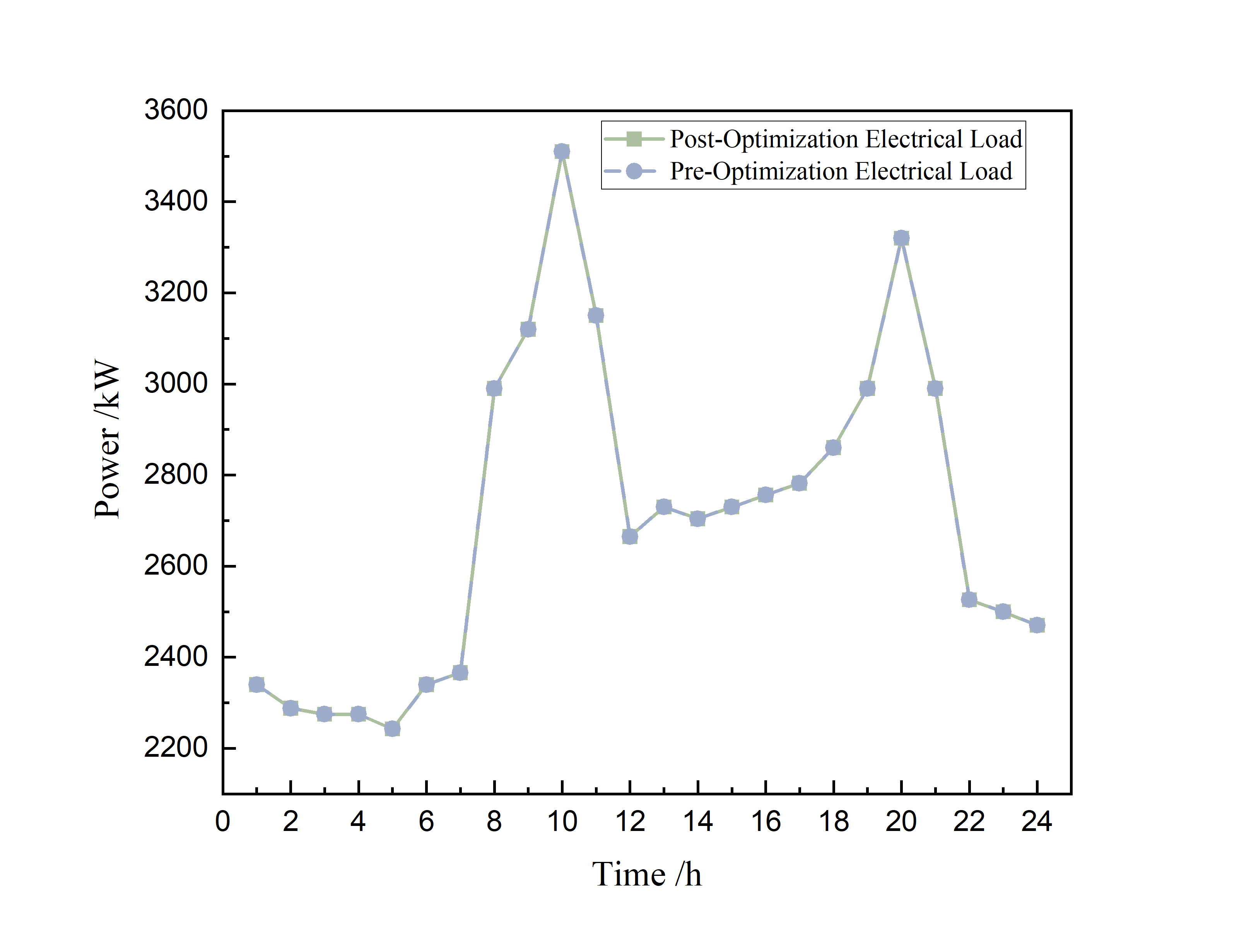}
\caption{Electric load curves before and after optimization in Scenario 3}\label{fig1}
\end{figure}

\begin{figure}[h]
\centering
\includegraphics[width=0.64\textwidth]{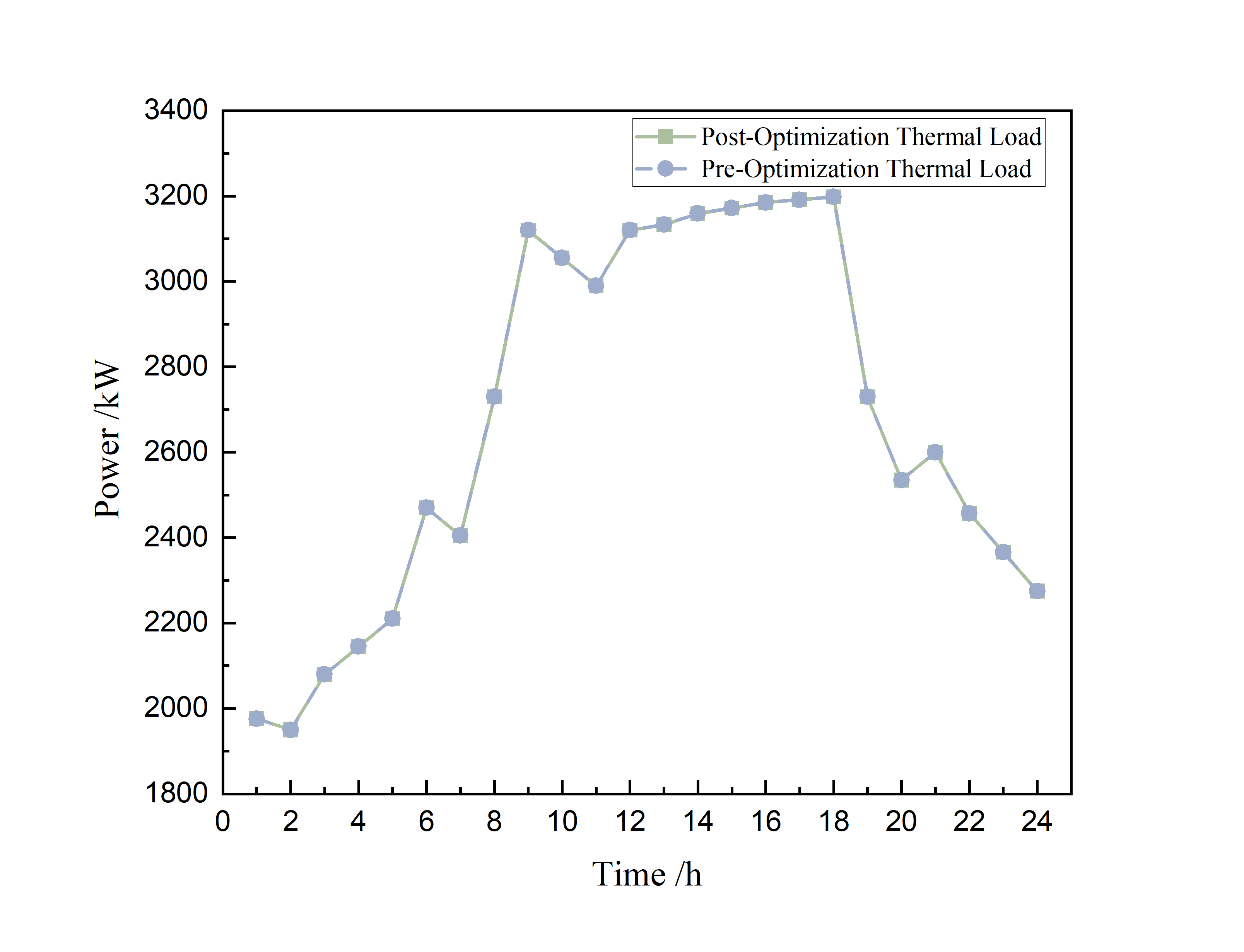}
\caption{Thermal load curves before and after optimization in Scenario 3}\label{fig1}
\end{figure}

\begin{figure}[h]
\raggedleft
\includegraphics[width=0.9\textwidth]{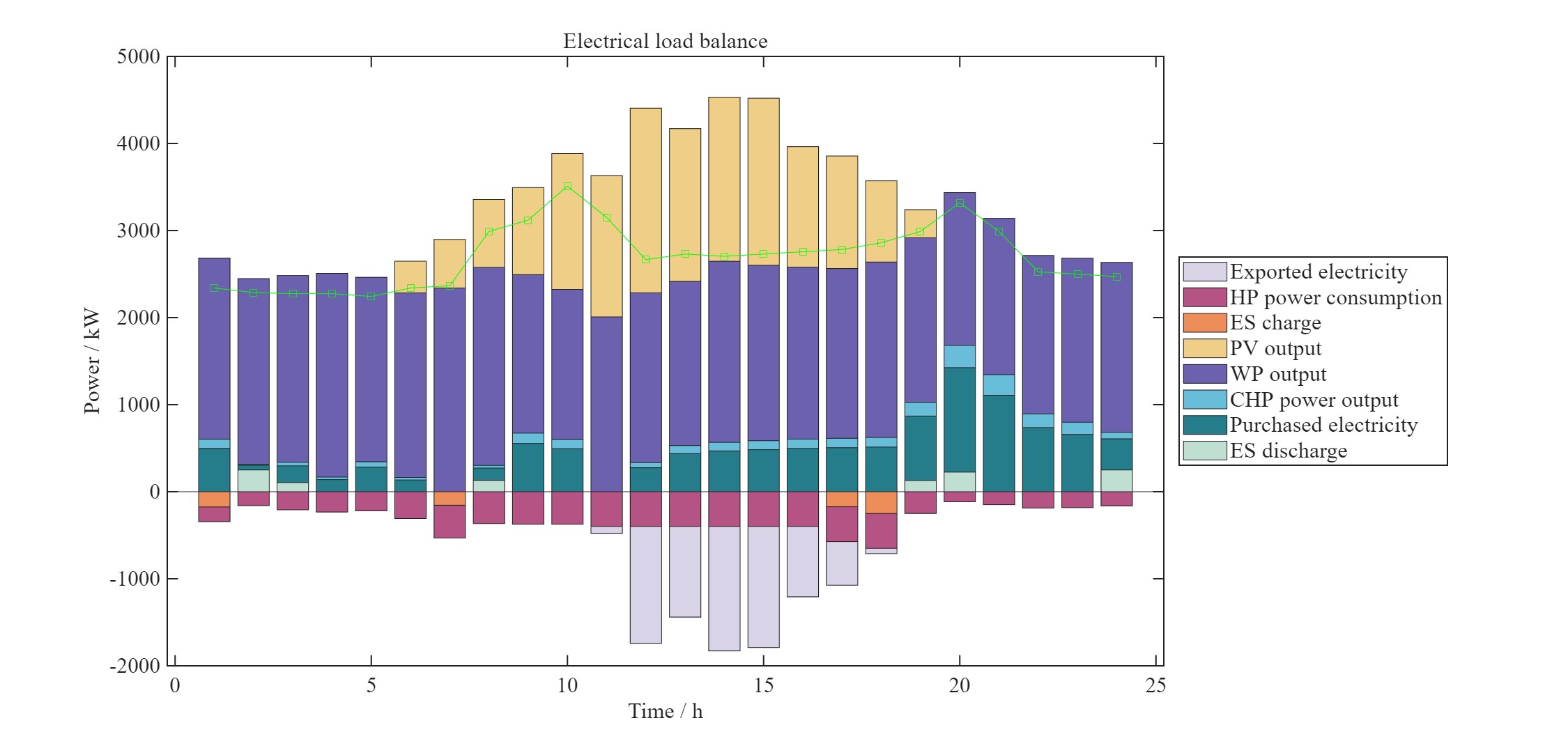}
\caption{Electric Load Balance Diagram of Scenario 3}\label{fig1}
\end{figure}

\begin{figure}[h]
\raggedleft
\includegraphics[width=0.9\textwidth]{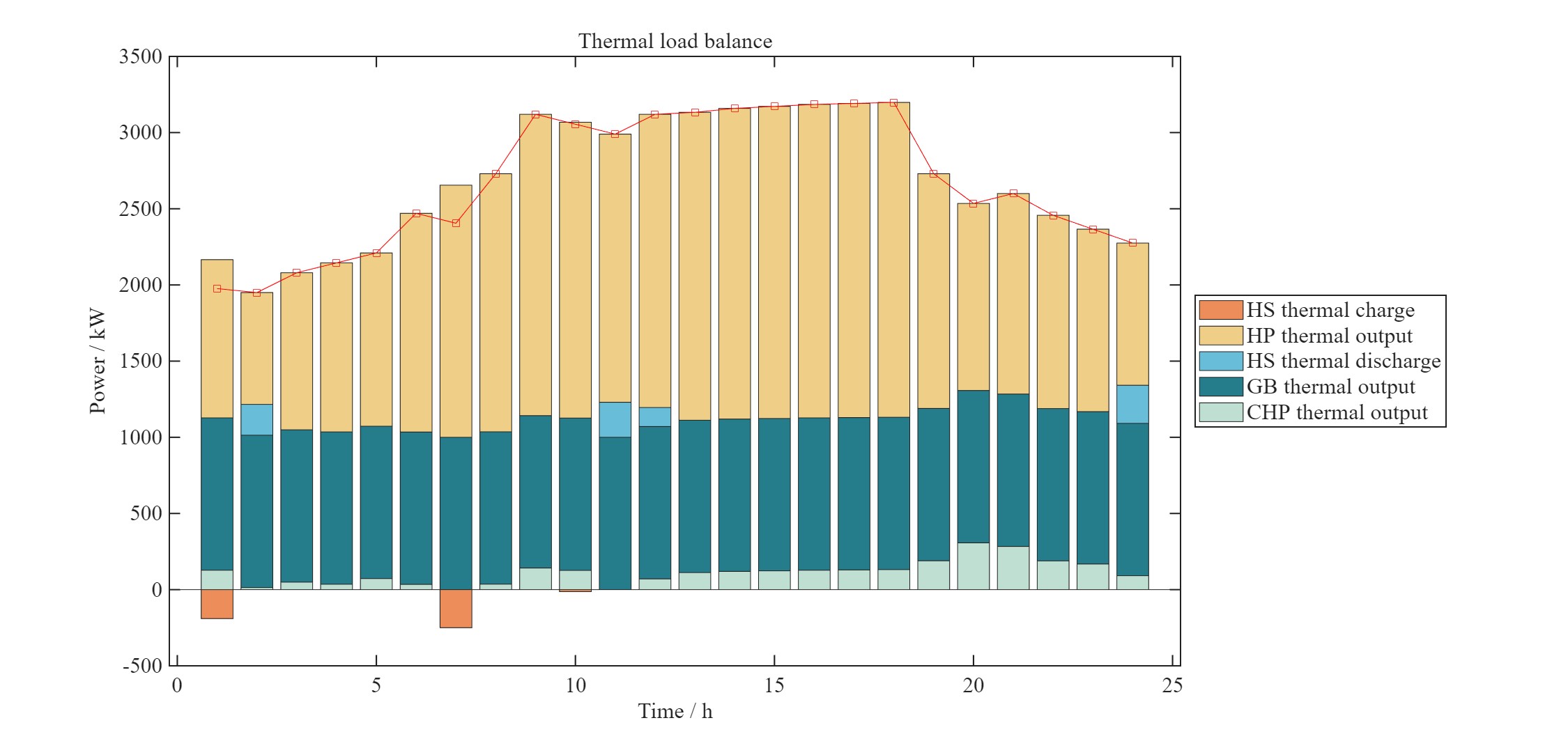}
\caption{Thermal Load Balance Diagram of Scenario 3}\label{fig1}
\end{figure}

\begin{figure}[h]
\centering
\includegraphics[width=0.64\textwidth]{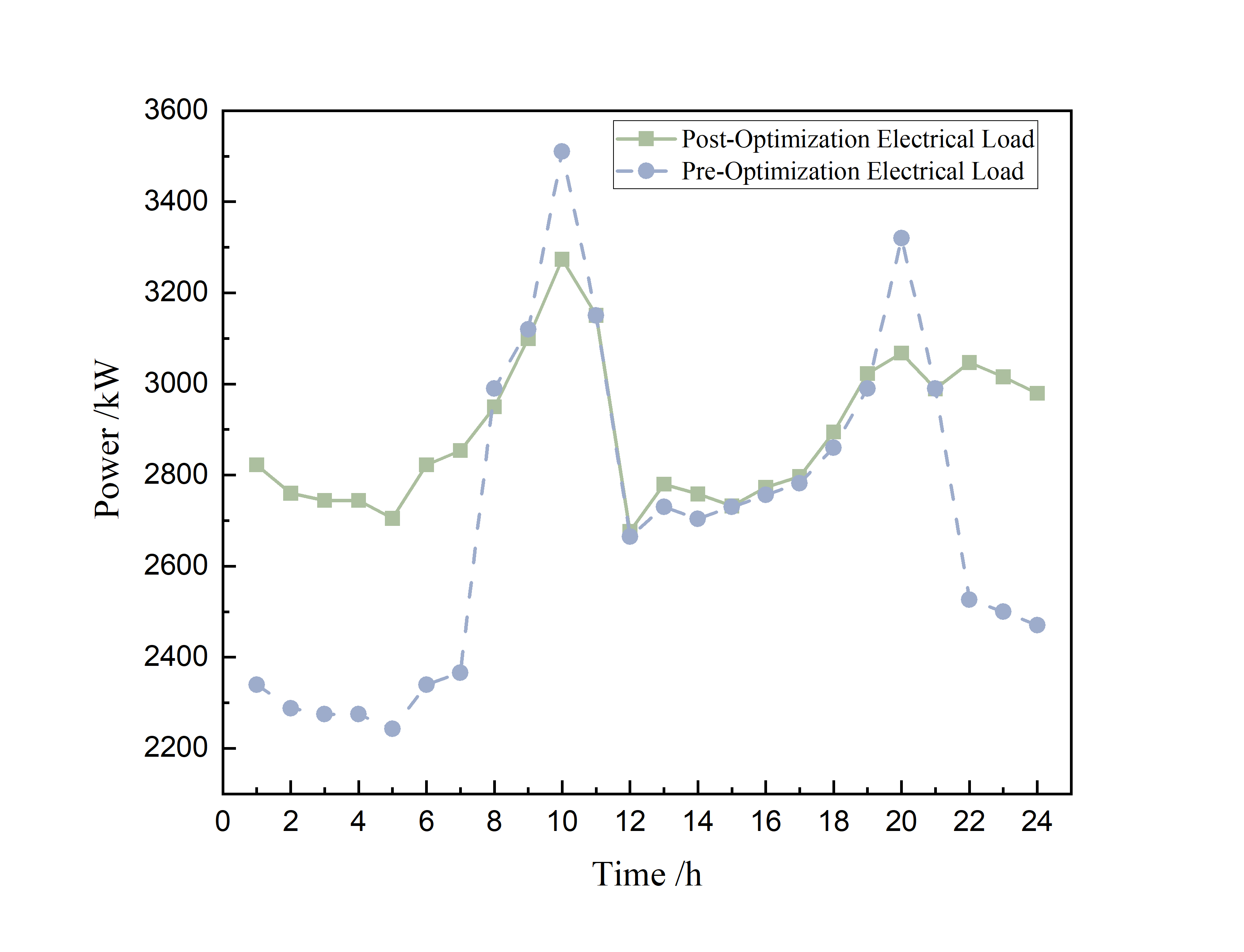}
\caption{Electric load curves before and after optimization in Scenario 4}\label{fig1}
\end{figure}

\begin{figure}[h]
\centering
\includegraphics[width=0.64\textwidth]{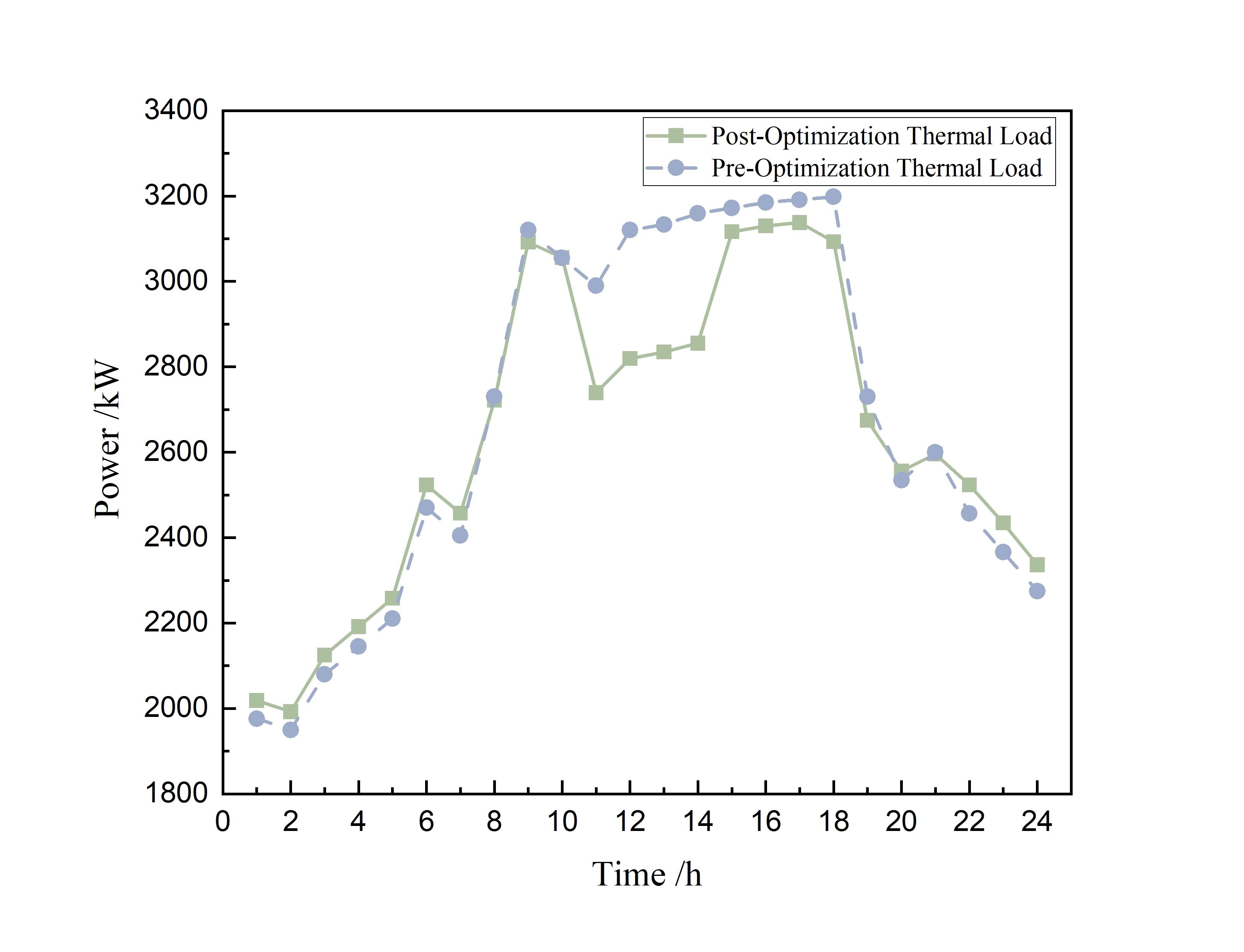}
\caption{Thermal load curves before and after optimization in Scenario 4}\label{fig1}
\end{figure}

\begin{figure}[h]
\raggedleft
\includegraphics[width=0.9\textwidth]{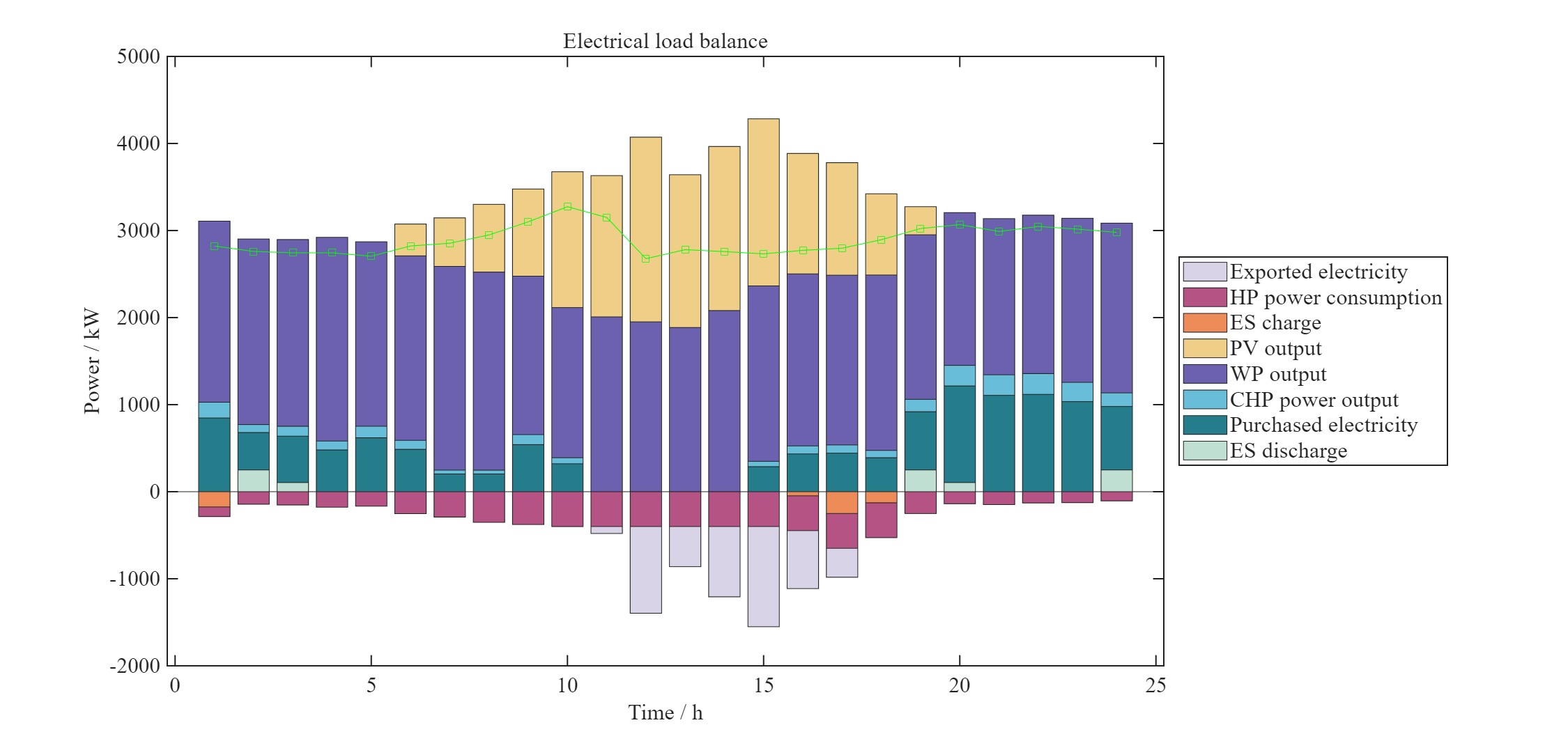}
\caption{Electric Load Balance Diagram of Scenario 4}\label{fig1}
\end{figure}

\begin{figure}[h]
\raggedleft
\includegraphics[width=0.9\textwidth]{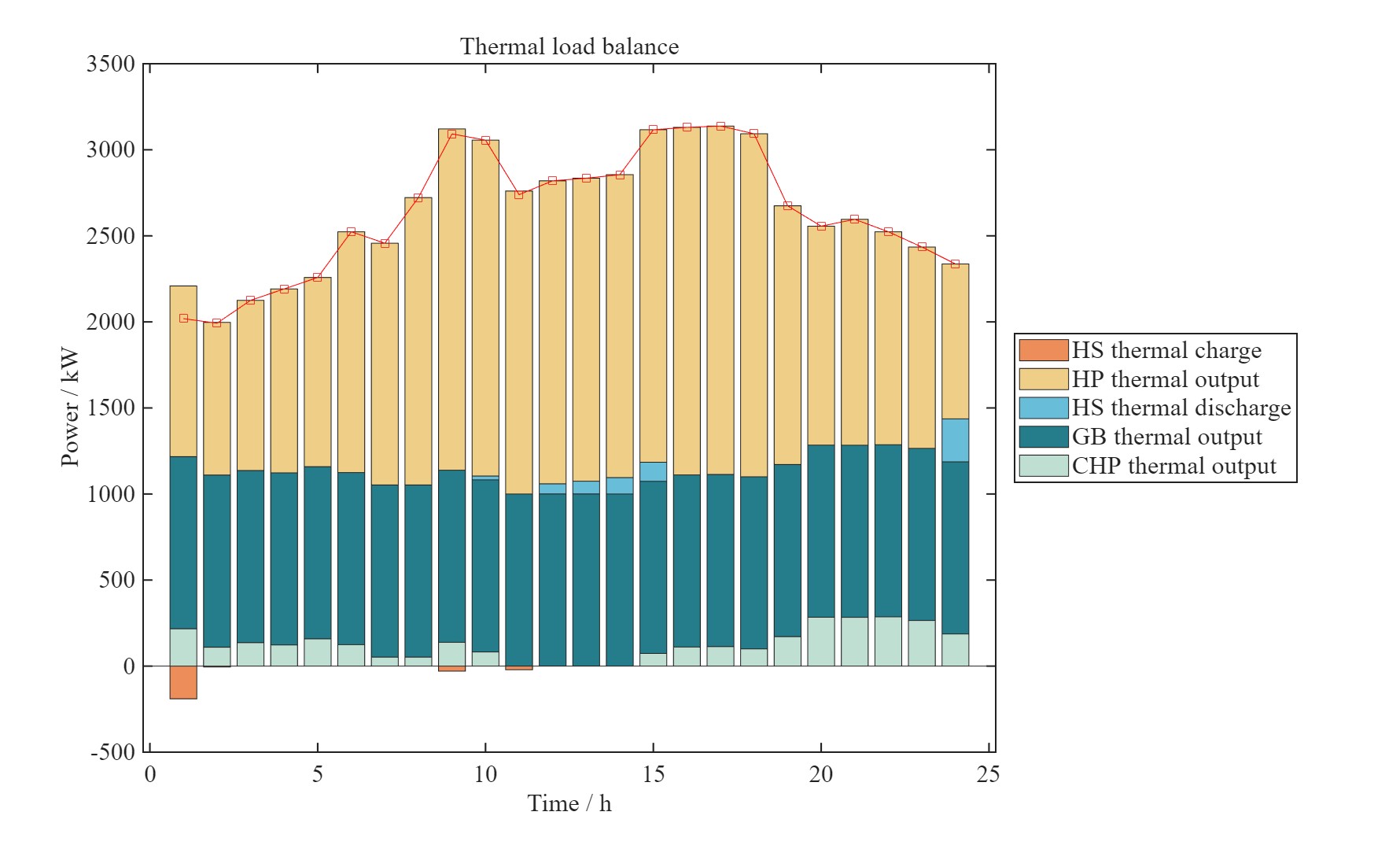}
\caption{Thermal Load Balance Diagram of Scenario 4}\label{fig1}
\end{figure}
Compared to the results for Scenario 3 shown in Fig.11 to 14, Scenario 4 demonstrates superior performance across multiple evaluation metrics, including total system operating costs, energy purchase expenditures, carbon trading fees, and total carbon emissions. This optimization is primarily due to the coordinated integration of the carbon trading mechanism and a dynamic DR strategy into the dispatch model. This strategy not only effectively guides users to proactively reduce or curtail electricity consumption during periods of high electricity prices and shifts some transferable load to periods of lower electricity prices, but also meets end-user energy needs through the substitution of thermal power, further smoothing the electricity load curve and improving system operational flexibility.\\

As shown in Fig.15 to 18, the load curves and load balance diagrams before and after the implementation of DR in Scenario 4 clearly demonstrate the temporal reconfiguration of load. Unlike the original load curve's distinct “peak-flat-valley” distribution pattern, after implementing the DR strategy, the system effectively shifts load from high-price periods to low-price periods, reducing load pressure during peak periods while simultaneously increasing energy usage during off-peak periods, smoothing the overall load curve. Furthermore, substitutable loads, guided by electricity price signals, enable conversion between energy forms: during peak electricity price periods, some demand previously met by electricity is converted to thermal energy. During low-price periods, the system converts thermal load back into electricity through equipment such as HP, optimizing energy allocation without compromising user comfort. The coordinated operation of price-responsive and substitutable DR mechanisms not only effectively controls system carbon emissions but also significantly improves the temporal distribution of load, achieving peak shaving and valley filling.\\

During off-peak electricity price periods, the system meets the power demand for HP operation, thermal storage charging, and electrical loads by combining WT and PV power generation with electricity purchased from the upper-level grid, thereby achieving power balance. This optimization strategy is based on the following comprehensive considerations: First, WP generation offers significant advantages, such as low operation and maintenance costs and renewable resources. Therefore, prioritizing wind energy resources during the scheduling process helps ensure energy supply reliability while effectively reducing the system's overall operating costs. Second, when WP output is insufficient to fully cover the power load, and if electricity prices are relatively low during that period, purchasing electricity from the upstream power grid is more economical than purchasing natural gas from the upstream gas source network for power generation or heating. Regarding the selection of heat energy supply paths, given the high thermal efficiency of HP systems, which significantly improve the heat output per unit of energy consumed compared to traditional GB, the system prioritizes HP equipment for efficient heating when meeting heat load demand. When the heat energy provided by the HP cannot fully cover the current heat load and the cogeneration unit is not in operation, the system will use the GB as an auxiliary heat source to ensure a balanced supply of heat load.\\

During flat-rate periods, the system primarily relies on the combined output of WP and CHP units to meet HP operation and power load requirements, while the heat load is supplied by both HP and CHP units. This strategy was developed based on an economic analysis: compared to off-peak periods, flat-rate periods have higher electricity purchase prices. During these periods, the cost of purchasing gas from the upstream gas grid is lower than the cost of purchasing electricity from the upstream grid, so the system's energy scheduling strategy was adjusted accordingly.\\

During peak electricity price periods, the system utilizes WP generation, CHP unit output, and HP discharge to meet HP operation and power load requirements. Heat load and thermal energy storage are simultaneously supplied by the HP and CHP unit. This strategy was chosen based on economic considerations: during peak electricity price periods, when electricity prices are higher, the cost of purchasing gas from the upstream gas grid is lower than the cost of purchasing electricity from the upstream grid. Therefore, the system prioritizes gas as a supplemental energy source.\\

In summary, this RIES's operational strategy fully incorporates the dynamic changes in electricity and natural gas procurement costs over time. It also comprehensively considers the operating efficiency and carbon emission characteristics of GT and GB, dynamically optimizing energy equipment output decisions. By selectively selecting low-carbon, efficient energy supply paths while ensuring economical system operation, the system not only effectively controls operating costs but also significantly reduces overall carbon emissions, achieving the goal of balanced economic and low-carbon development.\\
(2) Analysis of the carbon reduction potential of the carbon trading mechanism\\

In order to systematically evaluate the impact of the carbon trading mechanism on the carbon emission reduction effect and economic operation performance in the IES, this paper designed three typical comparison scenarios, namely Scenario 1, Scenario 3 and Scenario 4. Among them, Scenario 1 is a baseline scheme and does not consider the carbon trading mechanism and demand response strategy; Scenario 3 introduces the carbon trading mechanism on the basis of the baseline scenario to evaluate its independent role; and Scenario 4 further superimposes the dynamic DR strategy on this basis to form a situation of coordinated control of carbon trading and user-side response. By comparing the total operating cost and carbon emission level of the system under the above different scenarios, the emission reduction potential of the carbon trading mechanism and its comprehensive benefits brought by the synergy with other regulatory mechanisms can be effectively quantified.\\

By comparing the operating results of Scenario 3 and Scenario 1, it is not difficult to conclude that the carbon emission cost of Scenario 3 is 13.27\% lower than that of Scenario 1, and the actual carbon emissions are reduced by $4007.61\,\mathrm{kgCO_2}$. This result verifies that the carbon trading mechanism effectively alleviates the pressure of Scenario 1 to bear the full actual carbon emission cost by giving the system an initial carbon quota. At the same time, carbon quotas prompt the system to proactively optimize its energy usage strategy and reduce actual emissions, demonstrating the restraining effect of carbon trading prices on high-carbon emission behaviors.\\

A comparison of the operating results of Scenario 4 and Scenario 3 reveals that the former achieves significant reductions in key indicators such as total operating costs, carbon trading costs, energy purchase costs, and actual carbon emissions. This optimization achievement is primarily due to the combined benefits of the synergistic interaction between the carbon trading mechanism and the dynamic DR strategy. First, the system utilizes electric ES devices to charge during periods of low electricity prices and discharge during peak periods. Combined with the reverse regulation of thermal energy storage devices across different electricity price cycles (i.e., storing heat during low-price periods and releasing it during high-price periods), this effectively shifts and reduces loads over time, thereby smoothing the load curve and improving energy efficiency. Second, the mutual substitution mechanism between electricity and heat energy is utilized during the load response process, with electric-driven devices (such as heat pumps) replacing gas boilers to meet some of the heat load, reducing reliance on fossil fuels and further reducing total carbon emissions. In addition, the carbon trading mechanism allocates carbon emission quotas and allows the system to offset the cost pressure caused by excess emissions through market mechanisms, thereby achieving effective regulation of carbon emissions while ensuring the economic efficiency of the system.\\

The case study results for Scenario 4 demonstrate that the coordinated scheduling of electric-thermal ES significantly enhances the flexibility of this RIES park. Electrical ES reduces energy procurement costs through arbitrage of peak and off-peak electricity prices, while simultaneously decreasing fossil fuel consumption during high-carbon periods. Thermal ES utilizes renewable energy or low-cost electricity during low-price periods for heat storage, thereby replacing the direct operation of GB and further reducing carbon emissions. Compared with Scenario 3, Scenario 4 achieves a 2.56\% reduction in carbon trading costs, a 4.02\% decrease in total operating costs, and an actual carbon emission reduction of 1895.95 kg, illustrating the effectiveness of the synergistic strategy between carbon trading and dynamic DR in cost and carbon reduction. This coordination mechanism not only optimizes energy utilization efficiency but also enhances the system’s ability to cope with carbon price fluctuations through flexible dispatch. Given the advantages of Scenario 4 in carbon reduction and across different parameters, subsequent analysis of carbon reduction potential will focus on Scenario 4 to develop a “Carbon Emission–Carbon Sensitivity Factor Relationship Model" quantifying the impact of carbon sensitivity factors on carbon emissions, and evaluating the driving effect of carbon sensitivity factors on carbon emission reduction in RIES.\\ 

\subsection{Carbon Sensitivity Factor Analysis in RIES}\label{subsec2}
\noindent\hspace{1em}In RIES, various model parameters serve as key weights in the objective function of the optimization model. Variations in their values directly influence decisions regarding carbon emission control, carbon trading expenditures, and energy procurement strategies, thereby significantly impacting the overall carbon emission level of the system. To systematically evaluate the impact of different parameter variations on carbon emissions, this paper further conducts a parameter sensitivity analysis of the RIES, aiming to reveal the sensitivity of each parameter to the system's carbon reduction effectiveness.\\

A total of 27 RIES parameters are selected for analysis, comprising 24 system operational variables and 3 carbon market-related variables. The value ranges for these parameters are primarily determined based on current policy standards, equipment technical specifications, and typical operating conditions. The specific value ranges, carbon sensitivity levels, and Pearson correlation coefficients for each variable are presented in the Table 4.\\

The main steps of the sensitivity analysis are as follows: First, within the scenario considering the carbon market and electro-thermal demand response, the value ranges of the 27 parameters are normalized. Subsequently, the value of each parameter is individually varied within its dynamic adjustment range, and the corresponding system carbon dioxide emissions are calculated. By comparing the differences in carbon emissions caused by variations in different parameters, the system's carbon sensitivity response to each parameter is obtained.\\
\begin{figure}[h]
\centering
\includegraphics[width=0.78\textwidth]{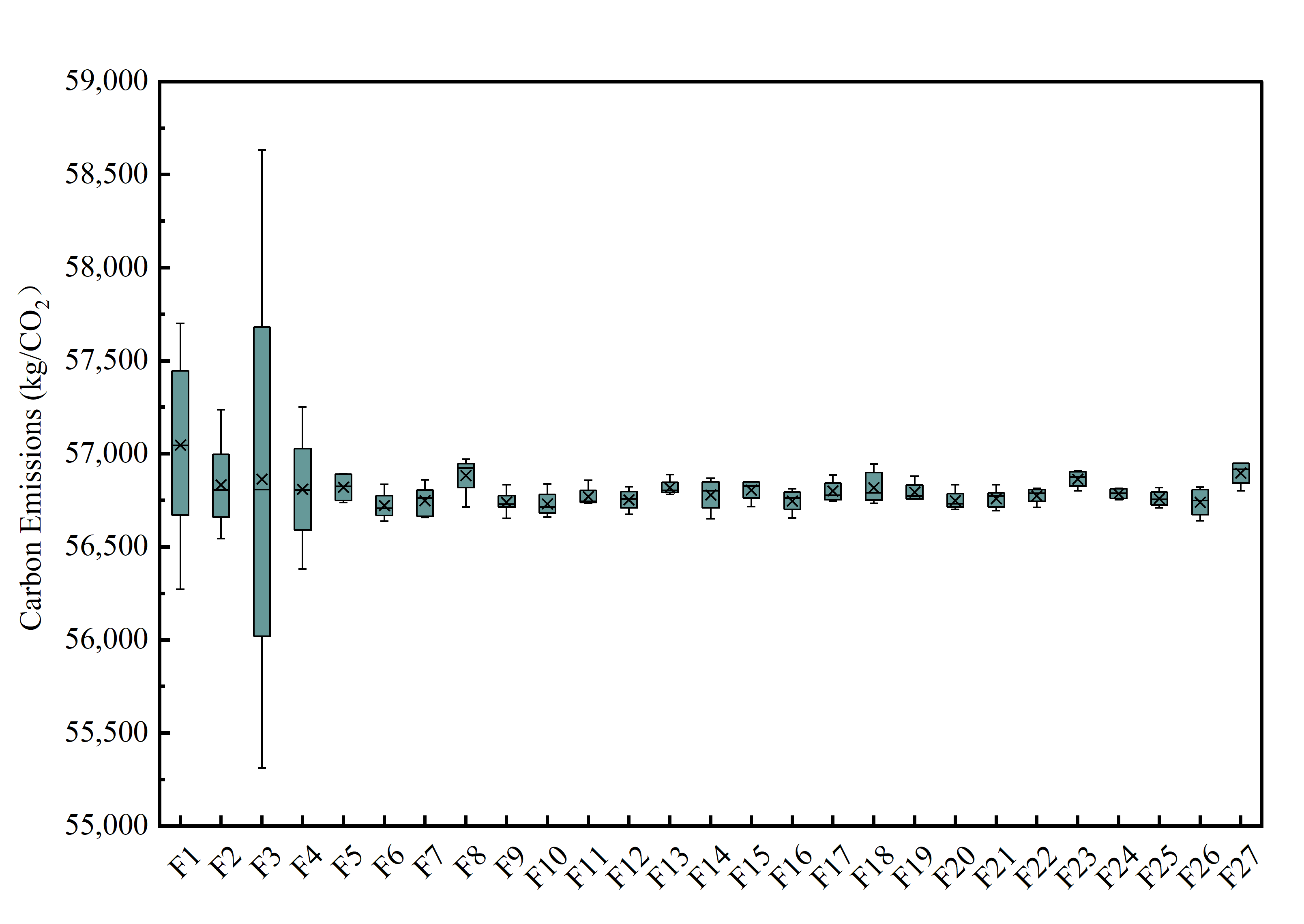}
\caption{Carbon sensitivity analysis of RIES parameters}\label{fig1}
\end{figure}

The analysis results, as illustrated in Fig. 19, indicate that the power generation efficiency of the GT, the heating efficiency of the gas boiler, the efficiency of the pyrolysis furnace, and the power generation efficiency of pyrolysis have the most significant impact on the carbon emissions of the RIES. Changes in these parameters play a dominant role in carbon reduction. The influence of the remaining parameters is relatively minor, exhibiting lower sensitivity concerning the system's carbon emissions.\\ 

To further investigate the impact mechanisms of different parameters on carbon emissions, this paper constructs a correlation analysis model between carbon emissions and the aforementioned key efficiency parameters. The calculated Pearson correlation coefficients, as shown in the table, indicate significant correlations between each efficiency parameter and carbon emissions.\\

As the values of the four high-sensitivity parameters increase, system carbon emissions show a clear downward trend, reflecting the positive contribution of efficiency improvements to carbon reduction. The specific reasons are as follows:\\

(1) Improvements in the power generation efficiency of the GT and the heating efficiency of the GB reduce primary energy consumption from the generation side. The increased efficiency of the GT reduces the losses during the conversion of fuel chemical energy to electrical energy, thereby lowering fuel consumption for the same power demand and avoiding redundant carbon emissions caused by low energy conversion efficiency. The improved heating efficiency of the gas boiler significantly enhances the conversion rate from fuel chemical energy to thermal energy, achieving the same heating effect with less fuel input. Compared to traditional inefficient heating methods in rural areas, such as raw coal or direct straw combustion (with thermal efficiencies of only 15\%–30\%), high-efficiency gas boilers (typically with efficiencies of 85\%–95\%), when coupled with clean gas, can substantially reduce fuel consumption per unit of heat output, thereby effectively cutting carbon emissions.\\

(2) Improvements in the efficiency of the pyrolysis furnace and pyrolysis power generation promote the clean and efficient utilization of biomass waste in RIES. Pyrolysis, a process that thermochemically converts biomass (e.g., straw, wood chips) into high-value-added products under anoxic conditions, benefits from efficiency gains by converting more biomass into energy products such as biochar, pyrolysis gas, and bio-oil, rather than losing it as ash or pollutants. Products like pyrolysis gas can be directly used for power generation or heating, substituting for fossil fuels. Simultaneously, the improvement in pyrolysis power generation efficiency reduces conversion losses across the entire energy chain from biomass to electricity, enabling more waste biomass resources (e.g., straw) to be converted into clean electricity that can replace coal-fired power, thereby further enhancing the system's overall carbon reduction performance.\\

Through correlation analysis of the 23 low-sensitivity factors, the reasons for their lower sensitivity can be categorized into three types, as shown in the table:\\

(1) Parameters exerting a direct influence on the carbon reduction process of RIES, exemplified by Parameters 4–11;\\

(2) Parameters directly associated with the carbon trading mechanism, which in turn indirectly affect the operation of RIES, including Parameters 13 and 14;\\

(3) Parameters directly governing the carbon reduction–oriented DR while indirectly shaping the energy mix through cost-weight modulation, such as Parameters 12 and 15–27.\\

\subsection{Recommendations for Carbon Reduction in RIES}\label{subsec2}
\noindent\hspace{1em}To achieve low-carbon and economic operation of RIES, this subsection proposes carbon reduction pathways from two perspectives: system operation mechanisms and parameter optimization. At the macro level, the synergistic optimization of the carbon trading mechanism and electro-thermal DR can enable active carbon reduction and cost optimization for the system, driven by market signals. At the micro level, based on parameter sensitivity analysis results, identifying and optimizing highly sensitive parameters that significantly impact carbon emissions can further enhance energy utilization efficiency and the level of clean energy substitution. The synergy between these two aspects can provide a systematic strategic basis for the long-term low-carbon operation of RIES.\\

(1) Recommendations for Carbon Reduction via Synergy between Carbon Market and Demand Response\\

A comparative study of four scenarios indicates that a synergistic optimization strategy incorporating carbon trading and dynamic DR offers significant advantages in selecting carbon reduction pathways for RIES. As shown in the case study analysis in Section 4.2, under the scenario considering the synergistic optimization of carbon trading and dynamic DR, the system's total operating cost, energy procurement cost, carbon trading cost, and carbon emissions all reach their minimum values. The carbon reduction achieved in this scenario is 3.23\% higher than that under the scenario with only the carbon market mechanism and 7.74\% higher than that under the scenario with only dynamic demand response. Therefore, it is recommended to prioritize the synergistic optimization strategy combining the carbon market mechanism and dynamic DR, leveraging carbon trading prices to guide user carbon reduction behaviors. The second recommended option is optimization operation considering only the carbon trading mechanism, followed by optimization operation considering only dynamic DR.\\

(2) Equipment-Level Carbon Reduction Strategies Driven by Carbon Parameter Optimization\\

The sensitivity analysis indicates that the carbon emissions of RIES are highly responsive to variations in multiple technical and economic parameters. Among them, efficiency-related parameters exert the most significant impact on system carbon emissions. Specifically, the efficiencies of GT power generation, GB heating, pyrolysis, and pyrolysis-based power generation exhibit a strong negative correlation with emissions. Quantitative analysis shows that a 1\% improvement in these efficiencies can reduce system carbon emissions by approximately 1.0\%–1.2\%. Therefore, enhancing the energy conversion efficiency of gas and biomass-coupled generation units is a priority for achieving substantial emission reductions in RIES.\\

In contrast, the average grid emission factor and system thermal loss rate show a positive correlation with carbon emissions, indicating that higher carbon intensity of purchased electricity or greater system heat losses increase overall emissions. In addition, the PV temperature coefficient exhibits a moderate positive correlation, suggesting that elevated ambient temperatures reduce PV output and consequently increase the system’s reliance on fossil electricity.\\

Moderately sensitive parameters mainly involve ES and waste heat recovery processes, including storage charge/discharge efficiency, waste heat recovery efficiency, and EB efficiency. Improvements of 5\% in these parameters can further reduce carbon emissions by approximately 2\%–4\%, reflecting their indirect contribution to emission reduction through enhanced energy utilization and reduced electricity purchases.\\

\begin{table}[htbp]
\caption{Carbon Sensitivity Parameters of RIES System Components}\label{tab:carbon_sensitivity}%
\scriptsize
\begin{tabular}{@{}p{0.35cm} p{0.8cm} p{2.8cm} p{1.2cm} p{1.3cm} p{5.8cm}@{}}
\toprule
No. & Symbol & Parameter & Sensitivity & Pearson & Theoretical Basis\\
\midrule
F1 & $\eta_{\mathrm{gt}}$ & Gas turbine electrical eff. & High & -0.9963 & Eq.9: Higher efficiency reduces natural gas use and CO$_2$. \\
F2 & $\eta_{\mathrm{gb},h}$ & Gas boiler thermal eff. & High & -0.9905 & Eq.10: Higher efficiency lowers fuel per heat. \\
F3 & $\eta_{\mathrm{pf}}$ & Pyrolysis gasifier eff. & High & -0.9985 & Eq.3: Higher efficiency increases biofuel, replacing fossil fuels. \\
F4 & $\eta_{\mathrm{pg}}$ & Pyrolysis power eff. & High & -0.9999 & Eq.7: Boosts renewable power share. \\
F5 & $\eta_{\mathrm{grid}}$ & Grid emission factor & Low & +0.55 & Higher emission factor increases CO$_2$ from grid imports. \\
F6 & $\eta_{\mathrm{loss}}$ & Thermal loss rate & Low & +0.33 & Higher losses require extra carbon‐intensive heat supply. \\
F7 & $\eta_{\mathrm{h}}$ & Waste heat recovery eff. & Low & -0.58 & Higher recovery reduces auxiliary heat demand. \\
F8 & $\eta_{\mathrm{st}}^{\mathrm{ch}}$ & Storage charging eff. & Low & -0.38 & Lower efficiency increases grid electricity use. \\
F9 & $\eta_{st}^{\mathrm{dis}}$ & Storage discharge eff. & Low & -0.42 & See No.~8. \\
F10 & $\eta_{\mathrm{eb}}$ & Electric boiler eff. & Low & -0.47 & Higher efficiency reduces electricity per heat unit. \\
F11 & $\gamma_{\mathrm{T}}$ & PV temp. coefficient & Low & +0.33 & Higher value decreases PV output at high temp. \\
F12 & $\eta_{\mathrm{B2G}}$ & Biogas→gas eff. & Low & -0.40 & Improves fossil substitution. \\
F13 & $e_{\mathrm{coal}}$ & Coal carbon allowance & Low & +0.45 & Higher allowance weakens decarbonization incentive. \\
F14 & $e_{\mathrm{gas}}$ & Gas carbon allowance & Low & +0.40 & See No.~13. \\
F15 & $c_{\mathrm{coal}}$ & Coal price coefficient & Low & -0.35 & Higher price shifts to lower‐carbon fuels. \\
F16 & $c_{\mathrm{gas}}$ & Gas price coefficient & Low & +0.32 & Higher price may shift back to coal. \\
F17 & $\beta_{\mathrm{ST}}$ & Wastewater coefficient & Low & +0.30 & Affects sludge mass and biogas potential. \\
F18 & $\eta_{\mathrm{AB}}$ & Fermentable organics frac. & Low & -0.38 & Higher fraction raises biogas production. \\
F19 & $\beta_{\mathrm{sludge}}$ & Sludge conversion coeff. & Low & +0.30 & Higher sludge raises treatment energy. \\
F20 & $\beta_{\mathrm{BG}}$ & Biogas production coeff. & Low & -0.40 & Higher conversion increases biogas. \\
F21 & $c_{\mathrm{curt}}$ & PV curtailment penalty & Low & +0.35 & Low penalty increases curtailment and fossil backup. \\
F22 & $\alpha_1$ & Inner wall heat transfer & Low & +0.20 & Poor insulation increases heat load. \\
F23 & $\alpha_2$ & Outer wall heat transfer & Low & +0.18 & See No.~22. \\
F24 & $\theta_1$ & Wall thermal conductivity & Low & +0.15 & See No.~22. \\
F25 & $\theta_2$ & Insulation conductivity & Low & +0.15 & See No.~22. \\
F26 & $\eta_{\mathrm{BD}}$ & Digester thermal eff. & Low & -0.33 & Lower eff. raises auxiliary heat. \\
F27 & $\eta_{\mathrm{eq}}$ & Elec→heat conversion eff. & Low & -0.30 & See No.~26. \\
\botrule
\end{tabular}
\end{table}
\FloatBarrier
Low-sensitivity parameters primarily relate to renewable energy and biomass utilization, such as biogas-to-gas conversion coefficient, fermentable organic fraction, sludge conversion coefficient, and mixed gas production coefficient. Their sensitivity coefficients range from 0.3 to 0.4, indicating that they indirectly enhance fossil fuel substitution by affecting biomass conversion and methane production.\\

Economic and carbon market parameters, including coal and natural gas cost coefficients and carbon allowances, influence emissions mainly through operational optimization rather than direct physical processes. For instance, a 10\% increase in the coal cost coefficient can reduce system carbon emissions by approximately 2\%, whereas a similar increase in gas carbon allowance may lead to a 1\%–1.5\% rise in emissions. This demonstrates the guiding role of economic signals and carbon trading mechanisms in low-carbon dispatch.\\

Additionally, structural parameters related to the biogas digester, such as heat transfer coefficients, wall thermal conductivity, and electrical-to-thermal conversion efficiency, exhibit minor influence on system emissions (with an absolute correlation of no more than 0.33), primarily affecting local heat losses rather than overall energy pathways.\\

In summary, improving the energy conversion efficiency of key generation units, particularly gas turbines and biomass-based power systems, yields the highest emission reduction benefits. Optimizing energy storage, waste heat recovery, and energy management strategies can further lower system carbon intensity, while economic and structural parameters serve as auxiliary levers, enabling flexible control of emissions via price signals and carbon trading. Achieving a 10\% efficiency improvement in highly sensitive parameters is estimated to reduce overall system emissions by approximately 8\%–10\%, providing an effective pathway toward low-carbon and economically efficient operation of RIES.\\

\section{Conclusion}\label{sec3}
\noindent\hspace{1em}This paper focuses on the carbon reduction potential of RIES. At the macro level, three carbon reduction optimization operation scenarios were constructed based on carbon trading mechanisms, electric-thermal DR, and their synergistic operation. By establishing a bi-level optimization model that considers both the main grid and the rural demand side, a quantitative assessment of the total system operation cost, carbon trading cost, and carbon emissions under different carbon reduction optimization scenarios was achieved. At the micro level, carbon sensitivity analysis was conducted on key system parameters to quantify their impact on carbon reduction potential. The study ultimately leads to the following conclusions:\\

(1) A biogas-straw biomass energy model is established on the supply side, and an electric-thermal DR model is developed on the load side. This achieves low-carbon coordinated operation of multiple energy sources in rural areas, facilitating the low-carbon and economic operation of RIES.\\

(2) At the macro level, models of electric-thermal DR and a tiered carbon trading mechanism are introduced. The results indicate that the synergistic optimization strategy integrating carbon trading with dynamic DR performs optimally in the carbon reduction pathway, achieving the lowest total system operating cost, energy procurement cost, carbon trading cost, and carbon emissions. Compared to the scenario with a standalone carbon trading mechanism, these metrics decreased by 4.02\%, 4.64\%, 2.56\%, and 3.23\%, respectively; compared to the scenario with standalone dynamic demand response, they decreased by 12.33\%, 12.28\%, 19.58\%, and 7.74\%, respectively. Therefore, it is recommended to prioritize the synergistic optimization strategy combining carbon market mechanisms with electric-thermal DR, followed by the operational mode considering only the carbon trading mechanism, and lastly, the operational mode considering only electric-thermal DR.\\

(3) At the micro level, carbon sensitivity analysis of key system parameters identified that the power generation efficiency of GT, the heating efficiency of gas boilers, and the efficiency of pyrolysis furnaces and pyrolysis power generation are highly sensitive parameters with significant impacts on carbon emissions. The correlation coefficients between both high- and low-sensitivity parameters and carbon emissions were quantified. The carbon reduction potential of the RIES was systematically evaluated, leading to proposed carbon reduction recommendations.\\

Several aspects remain for future research. For instance, this paper has not considered DR guided by dynamic carbon emission signals, which will be a focus of our subsequent work.

\section*{References}
\begin{enumerate}[
    label={\arabic*.},          
    labelwidth=2em,             
    labelsep=0.5em,             
    leftmargin=!,               
    align=right,                
    itemindent=0em              
]
    \item Ma T, Zhang S, Xiao Y, et al. (2023) Costs and health benefits of the rural energy transition to carbon neutrality in China. Nature Communications 14(1):6101.
    \item Peng L, Sun N, Jiang Z, et al. (2025) The impact of urban–rural integration on carbon emissions of rural household energy consumption: evidence from China. Environmental Development and Sustainability 27(1):1799–1827.
    \item Zhu Y, He Y, Zhou Q, et al. (2025) Impact of multi-energy complementary system on carbon emissions: insights from a rural building in Shangluo City, China. Energy Conversion and Management 327:119595.
    \item Li Y, Yang X, Du E, et al. (2024) A review on carbon emission accounting approaches for the electricity power industry. Applied Energy 359:122681.
    \item Zhao C, Wang J, Dong K, et al. (2023) A blessing or a curse? Can digital economy development narrow carbon inequality in China? Carbon Neutrality 2(15):1–24.
    \item Chen Q, Mao X, Hu F, et al. (2023) Optimal city carbon emissions in China from a theoretical perspective. Carbon Neutrality 2(1):32.
    \item Luo T, Shen B, Mei Z, et al. (2024) Unlocking the potential of biogas systems for energy production and climate solutions in rural communities. Nature Communications 15(1):5900.
    \item Shen G, Ru M, Du W, et al. (2019) Impacts of air pollutants from rural Chinese households under the rapid residential energy transition. Nature Communications 10(1):3405.
    \item Zhu Z, Qu Z, Gong J, et al. (2025) Robust optimal model for rural integrated energy system incorporating biomass waste utilization and power-to-gas coupling unit considering deep learning-based air conditioning load personalized demand response. Energy 321:135484.
    \item Khan SA, Tao Z, Agyekum EB, et al. (2025) Sustainable rural electrification: energy-economic feasibility analysis of autonomous hydrogen-based hybrid energy system. International Journal of Hydrogen Energy 141:460–473.
    \item Sánchez-Lozano D, Aguado R, Escámez A, et al. (2025) Techno-economic assessment of a hybrid PV-assisted biomass gasification CCHP plant for electrification of a rural area in the Savannah region of Ghana. Applied Energy 377:124446.
    \item Fu X, Zhou Y (2022) Collaborative optimization of PV greenhouses and clean energy systems in rural areas. IEEE Transactions on Sustainable Energy 14(1):642–656.
    \item Wu C (2025) Carbon emission evaluation and low carbon economy optimization scheduling of rural integrated energy system based on LCA method. IEEE Access 13:17182–17194.
    \item Gao Y, Zhang Y, Yun C, et al. (2025) Online optimization of integrated energy systems based on deep learning predictive control. Electric Power Systems Research 243:111510.
    \item Yang X, Wu C, Wang Z, et al. (2025) A two-stage dynamic planning for rural hybrid renewable energy systems under coupled carbon-green certificate trading. Energy 316:134336.
    \item Ma T, Zhang S, Xiao Y, et al. (2023) Costs and health benefits of the rural energy transition to carbon neutrality in China. Nature Communications 14(1):6101.
    \item Ju L, Wang P, Li Q, et al. (2022) Nearly-zero carbon optimal operation model and benefit allocation strategy for a novel virtual power plant using carbon capture, power-to-gas, and waste incineration power in rural areas. Applied Energy 310:118618.
    \item Jiang D, Gao H, He S, et al. (2024) A negative-carbon planning method for agricultural rural industrial park integrated energy system considering biomass energy and modern agricultural facilities. Journal of Cleaner Production 479:143837.
    \item Wang Y, Cai C, Liu C, et al. (2022) Planning research on rural integrated energy system based on coupled utilization of biomass-solar energy resources. Sustainable Energy Technologies and Assessments 53:102416.
    \item Sheng X, Lin S, Liang W, et al. (2025) Optimal long-term planning of CCUS and carbon trading mechanism in offshore-onshore integrated energy system. Applied Energy 379:124983.
    \item Zhao Y, Li M, Long R, et al. (2023) Techno-economic analysis of converting low-grade heat into electricity and hydrogen. Carbon Neutrality 2(19):1–22.
    \item Zhang Y, He Y (2025) Optimizing integrated energy systems: a multi-scenario scheduling approach with stepped carbon trading and two-stage hydrogen production under uncertainty. Applied Energy 401:126704.
    \item Tan J, Pan W, Li Y, et al. (2023) Energy-sharing operation strategy of multi-district integrated energy systems considering carbon and renewable energy certificate trading. Applied Energy 339:120835.
    \item Liu Y, Liu C, Yu W, et al. (2026) A two-tier optimization framework for urban integrated energy systems incorporating PSO-LSTM data-driven prediction and low-carbon demand response. Applied Energy 402:126967.
    \item Du B, Jia H, Zhang B, et al. (2025) Evaluating the carbon-emission reduction potential of employing low-carbon demand response to guide electric-vehicle charging: a Chinese case study. Applied Energy 397:126196.
    \item Sun X, Shi L, Zhang M, et al. (2024) Efficient and flexible thermal-integrated pumped thermal energy storage through composition adjustment. Carbon Neutrality 3(11):1–28.
    \item Cao L, Hu P, Li X, et al. (2023) Digital technologies for net-zero energy transition: a preliminary study. Carbon Neutrality 2(1):7.
    \item Ma X, Wu Y, Du Y, et al. (2024) Comprehensive performance evaluation of steam generating heat pumps for industrial waste heat recovery. Carbon Neutrality 3(35):1–26.
    \item Zhang C, Kuang Y (2024) Low-carbon economy optimization of integrated energy system considering electric vehicles charging mode and multi-energy coupling. IEEE Transactions on Power Systems 39(2):3649–3660.
    \item Wen C, Steadman S, Rafaq MS, et al. (2025) Can reduction of local carbon emissions motivate participation in demand-side flexibility programs? Evidence from the United Kingdom. Applied Energy 388:125610.
    \item Nowak C, Bertsch V (2025) Emission-based demand response in energy system optimisations—A systematic literature review. Applied Energy 401:126635.
    \item Yang P, Jiang H, Liu C, et al. (2023) Coordinated optimization scheduling operation of integrated energy system considering demand response and carbon trading mechanism. International Journal of Electrical Power and Energy Systems 147:108902.
    \item Gao X, Wang S, Sun Y, et al. (2024) Low-carbon energy scheduling for integrated energy systems considering offshore wind power hydrogen production and dynamic hydrogen doping strategy. Applied Energy 376:124194.
    \item Ma S, Liu H, Wang N, et al. (2024) Incentive-based integrated demand response with multi-energy time-varying carbon emission factors. Applied Energy 359:122763.
    \item Qin J, Pan X, Sun X, et al. (2026) Quantitative evaluation and sensitivity analysis of carbon emission reduction costs based on optimal scheduling of electric-thermal integrated energy systems. Electric Power Systems Research 250:112108.
    \item Wu D, Zhang T, Zhang J, et al. (2024) Sensitivity analysis and multiobjective optimization for rural house retrofitting considering construction and occupant behavior uncertainty: a case study of Jiaxian, China. Applied Energy 360:122835.
    \item Sahoo B, Debnath BK (2025) An integrated Pythagorean fuzzy group decision-making framework for sustainable biomass power plant site selection in rural communities. Energy for Sustainable Development 89:101868.
    \item Deng Y, Peng J, Ran X, et al. (2025) Energy, exergy, exergoeconomic and exergoenvironmental analyses of biomass heating systems for rural households: a case study in Northeastern China. Energy 333:137498.
    \item Chen Y, Yang Y, Liu X, et al. (2024) O2-H2O2 high-efficient co-oxidation of carbohydrate biomass to formic acid via Co3O4/C nanocatalyst. Carbon Neutrality 3(1):23.
    \item Yan L, Wang G, Xiang D, et al. (2024) Reductive amination of bio-platform molecules to nitrogen-containing chemicals. Carbon Neutrality 3(1):24.
    \item Xu W, Yu B, Song Q, et al. (2022) Economic and low-carbon-oriented distribution network planning considering the uncertainties of photovoltaic generation and load demand to achieve their reliability. Energies 15(24):9639.
    \item Yang H, Li C, Huang R, et al. (2023) Bi-level energy trading model incorporating large-scale biogas plant and demand response aggregator. Journal of Modern Power Systems and Clean Energy 11(2):567–578.
    \item Fu X, Wei Z, Sun H, et al. (2024) Agri-energy-environment synergy-based distributed energy planning in rural areas. IEEE Transactions on Smart Grid 15(4):3722–3738.
    \item Gao J, Meng Q, Liu J, et al. (2024) Multi-energy cooperative optimal scheduling of rural virtual power plant considering flexible dual-response of supply and demand and wind-photovoltaic uncertainty. Energy Conversion and Management 320:118990.
    \item Zhou T, Luo X, Liu X, et al. (2025) Quantification of rural residents’ willingness for electricity demand response: insights from field simulation experiments. Applied Energy 400:126589.
    \item Li DW, Huang JL, Yu D, et al. (2024) Development of low-carbon technologies in China’s integrated hydrogen supply and power system. Advances in Climate Change Research 15(5):936–947.
    \item Dong Z, Qing Z, Yu Z, et al. (2024) Performance response analysis and optimization for integrated renewable energy systems using biomass and heat pumps: a multi-objective approach. Carbon Neutrality 3(1):33.
    \item Sun X, Xie H, Qiu D, et al. (2025) Learning the reluctance of demand-side resources from equilibrium in price-based demand response. IEEE Transactions on Smart Grid 16(3):2699–2702.
    \item Li H, Hu H, Wu Z, et al. (2025) Modified predicted mean vote models for human thermal comfort: an ASHRAE database-based evaluation. Renewable and Sustainable Energy Reviews 209:115042.
    \item Yan Z, Li C, Yao Y, et al. (2023) Bi-level carbon trading model on demand side for integrated electricity-gas system. IEEE Transactions on Smart Grid 14(4):2681–2696.
    \item Lu Z, Bai L, Wang J, et al. (2023) Peer-to-peer joint electricity and carbon trading based on carbon-aware distribution locational marginal pricing. IEEE Transactions on Power Systems 38(1):835–852.
    \item Sohail A (2023) Genetic algorithms in the fields of artificial intelligence and data sciences. Annals of Data Science 10(4):1007–1018.
    \item Alhijawi B, Awajan A (2024) Genetic algorithms: Theory, genetic operators, solutions, and applications. Evolutionary Intelligence 17(3):1245–1256.
\end{enumerate}






\end{document}